\documentclass[12pt,preprint]{aastex}
\usepackage{longtable}
\usepackage{color}
\everymath{\rm}
\everydisplay{\sf}

\begin{document}

\title{Co-spatial Long-slit UV/Optical Spectra of Ten Galactic Planetary Nebulae with \textit{HST}/STIS 
II. Nebular Models, Central Star Properties and He+CNO Synthesis\altaffilmark{1}}
\author{
R.B.C. Henry\altaffilmark{2},
B. Balick\altaffilmark{3}, 
R.J. Dufour\altaffilmark{4}, 
K.B. Kwitter\altaffilmark{5}, 
R.A. Shaw\altaffilmark{6},
T.R. Miller\altaffilmark{2},\\
J.F. Buell\altaffilmark{7} 
and Romano L.\ M. Corradi\altaffilmark{8,9}
}

\altaffiltext{1}{Based on observations with the NASA/ESA \textit{Hubble Space Telescope} obtained at the Space Telescope Science Institute, which is operated by the Association of Universities for Research in Astronomy, Incorporated, under NASA contract NAS5-26555.}
\altaffiltext{2}{Department of Physics and Astronomy, University of Oklahoma, Norman, OK 73019}
\altaffiltext{3}{Department of Astronomy, University of Washington, Seattle, WA 98195}
\altaffiltext{4}{Department of Space Physics and Astronomy, Rice University, Houston, TX  77251}
\altaffiltext{5}{Department of Astronomy, Williams College, Williamstown, MA  01267}
\altaffiltext{6}{National Optical Astronomy Observatory, Tucson, AZ  85719}
\altaffiltext{7}{SUNY College of Technology at Alfred, Alfred, NY 14843}
\altaffiltext{8}{Instituto de Astrof{\'{\i}}sica de Canarias, E-38200 La Laguna, Tenerife, Spain}
\altaffiltext{9}{Departamento de Astrof{\'{\i}}sica, Universidad de La Laguna, E-38206 La Laguna, Tenerife, Spain}


\begin{abstract}

The goal of the present study is twofold. First, we employ new HST/STIS spectra and photoionization modeling techniques to determine the progenitor masses of eight planetary nebulae (IC~2165, IC~3568, NGC~2440, NGC~3242, NGC~5315, NGC~5882, NGC~7662 and PB6). {Second, for the first time we are able to compare each object's observed nebular abundances of helium, carbon and nitrogen with abundance predictions of these same elements by a stellar model that is consistent with each object's progenitor mass.} Important results include the following: 1)~the mass range of our objects' central stars matches well with the mass distribution of other PN central stars and white dwarfs; 2)~He/H is above solar in all of our objects, in most cases likely due to the predicted effects of first dredge up; 3)~most of our objects show negligible C enrichment, probably because their low masses preclude 3rd dredge-up; 4)~C/O versus O/H for our objects appears to be inversely correlated, {perhaps} consistent with the conclusion of theorists that the extent of atmospheric carbon enrichment from first dredge-up is sensitive to a parameter whose value increases as metallicity declines; 5)~stellar model predictions of nebular C and N enrichment are consistent with observed abundances for progenitor star masses $\le$1.5~M$_{\odot}$. Finally, we present the first published photoionization models of NGC~5315 and NGC~5882.

\end{abstract}

\keywords{galaxies: abundances, galaxies: evolution, ISM: abundances, planetary nebulae: general, stars: evolution }

\section{Introduction} 

Planetary nebulae (PN) are long believed to be major sources of cosmic carbon and nitrogen.
There is strong evidence that the elements C and N appear to be synthesized and ejected by both massive stars (M$>$8M$_{\odot}$) and low and intermediate mass stars (0.8M$_{\odot}$$\le$M$\le$8M$_{\odot}$; LIMS). This idea is supported by the existence in the Milky Way and other galaxies of carbon stars and C- and/or N-rich planetary nebulae (PN), both with LIMS as progenitors, as well as WC and WN stars with progenitor masses exceeding 20~M$_{\odot}$.  Thus, both components of the mass spectrum appear to affect the chemical evolution of these two elements, although their proportional contributions are uncertain.

Numerous galactic evolution studies have attempted to sort this problem out by modeling the observed changes in the abundance ratios of C or N with respect to some fiducial element such as Fe or O. Or in the case of a specific galaxy, the ratios of C/O and N/O as functions of galactocentric distance are often used as model constraints. [See papers by \citet{chiappini97}, \citet{henry00b}, \citet{chiappini03}, \citet{carigi05} and \citet{gavilan05,gavilan06} for specific examples.] These models account for finite stellar lifetimes, which is particularly important when considering LIMS, since the time delay due to their protracted evolution means that their nuclear contributions to the environment lag behind those of more rapidly evolving massive stars. A review of the papers cited above suggests that the question of the stellar origin of C and N is unresolved. 

The detailed chemical evolution models referred to above require as part of their input stellar yields, i.e., the amount of a specific isotope that is produced and ejected by a star of a certain mass over its lifetime. Determination of these yields is a complex endeavor, as it necessitates the construction of a grid of detailed stellar evolution models, which take a {    star's} initial mass and metallicity and in some cases rotation rate into account.  These same models also must incorporate the physics of nuclear reactions and stellar mass loss, along with assumptions regarding mixing length theory. Constraining such models relies heavily on observations of abundances in galaxies as well as in the remnants of individual evolved stars, i.e., supernova remnants and planetary nebulae. In the current project we focus on constraints for the production of C and N by LIMS.

The main nuclear processes that generate C and N are reasonably well understood---the dominant $^{12}$C nucleus comes primarily from the triple-alpha reaction during helium burning, while $^{14}$N is synthesized via the CN and ON cycles during hydrogen burning. Temperatures of roughly 100~MK and 15 MK, respectively, are necessary for initiating these reactions. Such temperatures are achievable within the cores of LIMS as well as massive stars during quiescent burning stages, i.e., main sequence and horizontal branch. However, during the post main sequence stages of LIMS these temperatures also arise in the H- and He-burning shells, i.e., regions external to the stellar core which are rich in H and He, respectively. Thus, significant nuclear processing may occur which results in C and N production. Eventually the fresh C or N is dredged up via convective processes to the stellar surface whence it is ejected in an enriched wind. {For detailed reviews of this topic see \citet{marigo03}, \citet{groenewegen04}, \citet{lattanzio04}, \citet{herwig05}, and \citet{karakas14}.}

The total amounts of C and N as a function of stellar mass which are synthesized and ejected by LIMS, as well as the fraction of each ejected element that is partitioned into the resulting PN, have been predicted by {numerous theorists, e.g., \citet{buell97, marigo01, karakas10}.} While it is unclear how one would measure the total quantities of ejected C and N directly to check on these computations, it is possible to measure the masses of these elements contained within a PN which has formed around the dying star. Comparing the latter determinations with the model predictions allows us to gauge the relevance of the models, including the total element yields of C and N that they predict. 

{Previously, the comparison between observation and theory has been made by first plotting the observed abundances of element X versus element Y for a sample of PN and then in the same graph plotting model tracks from stellar evolution models for a range of birth masses and constant metallicity. The point is to see if the trends described by the observations and model tracks are consistent [cf. \citet[Fig.~3]{marigo03}]. However, an attempt to directly connect observed progenitor masses and PN abundances with their theoretical counterparts in order to directly check on the goodness of the models has not to our knowledge been made.}
 
The goal of the work described here is to directly compare for the first time the nebular abundances of C and N observed by \citet[hereafter, Paper~I]{dufour15} for eight PN with predictions of post-AGB models as a function of progenitor mass. (Data in Paper~I are new co-spatial HST/STIS observations, observed through a single slit, free of atmospheric seeing, dispersion, and reddening effects, spanning a wavelength range of 1150--10270\AA, of ten PN representing a range in N abundance but with overall metallicities close to solar to control for metallicity effects.) We accomplish this comparison by computing a photoionization model of each of our objects that matches the calibrated and extinction-corrected line fluxes and ratios in order to derive the central star properties. From these results we derive the birth mass of each progenitor, combine it with our C and N abundances from Paper~I, and compare with several sets of published stellar model predictions of PN CNO abundances, in particular those of \citet{karakas10}. Past comparisons by \citet{marigo03} and \citet{karakas09} have been done by plotting observed nebular abundances of PN along with model tracks of nebular abundance predictions in an abundance versus abundance graph (e.g., C/O vs. O/H). Here we directly compare the inferred progenitor masses coupled with the observed C and N abundances from Paper~I to the model predictions of nebular abundance for a specific progenitor mass and metallicity.

The PN which we consider here are IC~2165, IC~3568, NGC~2440, NGC~3242, NGC~5315, NGC~5882, NGC~7662 and PB6. To our knowledge, in the cases of NGC~5315 and NGC~5882, our photoionization models are the first ones to be published for these objects.

Section~2 discusses the procedure used for modeling each object along with the results in each case. In \S~3 we present central  and progenitor star properties as inferred from the best model of each PN. Section~4 contains a comparison between our inferred nebular abundances and progenitor star masses with the predictions of several sets of stellar evolution models in the literature. We present our conclusions in \S~5.

\section{The Photoionization Models}

\subsection{Modeling Procedure}

We computed photoionization models of eight of the 10 PN with the goal of determining important properties of each central star such as effective temperature, luminosity, mass and radius.  The objects which we chose to model were IC~2165, IC~3568, NGC~2440, NGC~3242, NGC~5315, NGC~5882, NGC~7662, and PB6. NGC~3242 is also the subject of a focused paper \citep{miller15}. NGC~6778 and NGC~6537, the remaining two members of the original group of 10 PN, were excluded from our analysis because of insufficient data.

The modeling was carried out using the photoionization code CLOUDY \citep{ferland13} version 13.02. Based upon a set of input parameter values, CLOUDY steps outward from the inner radius of the nebula through the gas as it computes both an ionization balance and energy balance at each point. The model is truncated at the distance from the central star where the predicted and observed line ratios of [O~II]/H$\beta$ closely match\footnote{{We decided on this approach after computing many models and finding that the only way we could obtain simultaneous agreement with important observational constraints on stellar effective temperature and luminosity, [O~III] electron temperature, and line ratios such as He~II $\lambda$4686/He~I $\lambda$5876 and [O~III] $\lambda$5007/[O~II] $\lambda$3727 was to truncate the nebula at the point where the observed and predicted strengths of $\lambda$3727 agreed. This recognizes the fact that O$^+$ is normally the dominant oxygen ion in the outer region of the ionized zone. By truncating the nebula and forcing it to be matter bounded, the predicted strength of $\lambda$3737 was forced to match its observed counterpart without jeopardizing the predicted values of the other important diagnostics cited.}}. At each location the program accounts for the effects of photoionization and recombination as well as electron-ion collisions. The main output contains the predicted line strengths of thousands of emission lines, including those which are important for PN studies. The predicted strengths were compared to their observed counterparts, and then one or more input parameter values were changed in an effort to improve agreement between the two in the subsequent model. This iterative process was continued until the root-mean-square (RMS) of the relative difference between predicted and observed values for {line intensities with respect to H${\beta}$ was reasonably consistent with the RMS of the relative observational uncertainties for the same set. In addition, diagnostic ratios such as those sensitive to temperature or density were also checked for consistency.}

Central star input parameters included its effective temperature and luminosity, as well as the spectral energy distribution, which in all cases in our study was a blackbody. Nebular input parameters included a radially constant gas density starting at an inner radius inside of which the nebular density was zero, a complete set of chemical abundances, and values for the filling factor and dust parameter, the last controlling the dust/gas ratio. {(We assumed the {\it planetary nebula} grain set internal to CLOUDY.) The filling factor and dust parameter were treated as free parameters and were used to ``fine tune'' the model once an approximate model had been obtained. We also assumed a static spherical geometry, while the effects of background cosmic rays and photoelectric heating were ignored. The inner radius of each model was set to $10^{17}$cm throughout the analysis. Finally, each model was the result of two iterations, where the second one used values from the first as estimates for certain nebular conditions. We discuss nebular morphology considerations in section~\ref{morph} below.}  

The particular line strengths used to constrain the models were selected from the complete line list presented in Tables 3 and 4 in Paper~I and are listed here in the left column of Tables~\ref{lsa} and \ref{lsb}. The lines chosen were normally the strongest ones of a particular ion.

Generally, to compute a model we first set the {total hydrogen} gas density, using either the C~III] or [S~II] density, or the density derived from the nebular H$\beta$ luminosity {as a guide. The C~III] density was always the preferred one, because the relevant emission lines originate from nebular locations well inside the cloud and not at the ionization boundary followed by the [S~II] density. But in cases in which neither of those two could be determined due to high or low density sensitivity limitations, we used the H$\beta$ density. In any case, the chosen hydrogen density merely served as a starting point. Then with each successive model the density was adjusted as needed to help bring the nebula's observed and predicted H$\beta$ luminosity into closer agreement. Since our models assumed a constant hydrogen density, it proved to be difficult to match all three density values simultaneously.} 

After determining an initial density, we then adjusted the central star temperature until the observed and model ratio of He~II $\lambda$4686/He~I $\lambda$5876  matched. Next, the stellar luminosity was adjusted until the predicted nebular H$\beta$ luminosity matched the observed value. The input elemental abundances were initially set to their observed values reported in Paper~I. Then in subsequent interations the total metallicity, {i.e., the sum of all metal abundances, was scaled up or down} in order to force the electron-temperature-sensitive ratio [O~III] $\lambda$4363/$\lambda$5007 into agreement with its observed value\footnote{{The metals included in our models were: C, N, O, Ne, S, Ar, Na, Mg, Al, Si, Ca, Fe and Ni}}. Abundances of individual metals were varied to improve agreement for strengths of lines associated with those metals. {Finally, general limitations to our models include the use of blackbody spectral energy distributions, the assumptions of constant density and absence of shock heating throughout each nebula and the fact that we based each model on a spatially integrated long slit spectrum.}  

\subsubsection{Morphological Considerations\label{morph}}

In this paper abundances were determined by computing photoionization models over ranges of input parameters until a good match to the measured line intensities and ratios was obtained.  The photoionization code constructs a one-dimensional nebular model with user-specified radial density distribution, an inner radius, a density (which we adopted from a range that is commensurate with our observations), and a criterion for stopping the model at a particular radius.  As noted earlier, we elected to stop the integration when the [O~II] $\lambda$3727/H$\beta$ flux ratio matched the observations.

This procedure can lead to possible systematic errors.  For example, a one-dimensional model may not be a good representation of nebulae that are knotty or complex in outline (e.g. NGC~2440, PB6, NGC~5315; see Fig. 1 of Paper I for target images and slit placements).  For another, the choices of inner radius, density, and stopping criteria all potentially affect the nebular abundances and stellar properties such as effective temperature and luminosity inferred from the model.  

In this section we summarize a few tests that illuminate the likely systematic errors related to our choices of the model parameters.  We selected NGC~3242 for further studies, since it is simple in structure and all of its emission lines are moderately bright.  A set of new line fluxes and ratios were computed using the photoionization code CLOUDY as we first increased the inner radius (by a factor of three), then decreased the density (by 60), and lastly applied both of these changes.  We then compared these outcomes to those of the corresponding best-fit (``baseline'') models reported earlier.  

For example, when we moved the inner radius a factor of three farther from the central star, the H$\beta$ luminosity dropped by 7\%. At the same time, [O~II] $\lambda$3727/H$\beta$, [N~II] $\lambda$6584/H$\beta$, and [S~II] $\lambda$(6717+6731)/H$\beta$ increased by a factor of 2. C~III] $\lambda$1907/H$\beta$, [Ne~III $\lambda$3869/H$\beta$, and [O~III] $\lambda$5007/H$\beta$ decreased by 19\% or more, while C~IV $\lambda$1549/H$\beta$ and He~I $\lambda$5876/H$\beta$ decreased by more than 50\% and the N~V $\lambda$1240 all but vanished.  These changes are all in the sense of increasing the disagreement between observations and models.  From this we conclude that our choice of inner nebular radius is well constrained in the baseline model.

Decreasing the density such that the Str\"{o}mgren radius became twice as large had very little effect on the H$\beta$ luminosity or the line ratios.  The most substantial changes are N~V $\lambda$1240/H$\beta$ (+136\%), C~IV $\lambda$1549/H$\beta$ (+30\%), He~I $\lambda$5876 (-30\%)/H$\beta$, and the He~II lines (+25\%)/H$\beta$.  All of the He line ratios departed unfavorably and substantially relative to the baseline model while the lines of C~IV and N~V] improved slightly.  We surmise that changes in the density have little overall impact.

Of course, other model parameters that we adopted could also be in error.  Of note is the far-UV stellar flux, since the baseline models consistently underestimate the C~IV/H$\beta$ and N~V/H$\beta$ ratios by large fractions.  In addition, using the adopted [O~II] $\lambda$3727 cutoff gives very low values for the [S~II] lines.  This is no surprise since much if not most of the S$^+$ lies beyond the edge of the Str\"{o}mgren sphere.  That is to say, running models in which the [S~II]/H$\beta$  line ratios are fitted well results in predicted values for [O~II]/H$\beta$  and [N~II]/H$\beta$  that are much too large compared to the observations.

Most of our nebulae are round or mildly elliptical in geometry.  Thus the size of the inner nebular radius varies somewhat with slit position angle, so models along a single slit position may introduce errors in our abundance analysis.  However, we feel that the related systematic  uncertainties are small, since the variations in the actual inner radii are far smaller than the factor of 3 discussed above.  In addition, narrow-band images show that the ionization structure of elliptical PN is much the same at every position angle that avoids small knots embedded inside the nebula.  Moreover, our slit positions were not always placed radially through the nebular center.  The bottom line is this: We presume (but admittedly have not fully demonstrated) that computing photoionization models over a range of inner radii corresponding to various position angles in elliptical PN, and averaging the abundances in each position angle, is not warranted.  The detailed analysis of the present data for NGC~3242 in preparation \citep{miller15} will explore this question more thoroughly.

NGC~2440 is the only bipolar nebula in this sample.  We did not attempt to assess the effects of highly-bipolar morphology on the line ratios and abundances.  The challenge of modeling the detailed ionization structure of bipolar objects is daunting, involving 16 distinct morphological parameters, e.g. \citet{wright11}, and lies far beyond the scope of this paper.  Had our slits been placed along the symmetry axes of NGC~2440 then the photoionization models would probably have been sufficient to find nebular abundances (since the radiative transfer on that axis is very nearly one-dimensional in nature).  However, our slit was placed obliquely to this axis, such that it passed through zones of knotty and highly variable ionization.

\subsection{Model Results}

\subsubsection{General Analysis}

The model results for each of the eight PNe are given in Tables~\ref{lsa} and \ref{lsb}. As noted above, column~1 of each table contains the list of observed lines used to constrain the models. Note that compared to the more extensive list provided in Tables~3 and 4 of Paper~I, this line list contains only the lines that are crucial for constraining important stellar and nebular parameters. Each set of three columns that follows shows the observed and model line strengths relative to H$\beta$=100 along with the ratio of those two quantities, i.e., model/observed. We shall discuss the model for each object separately below. However, we note that generally the models successfully match a vast majority of the emission lines, as indicated by the preponderance of model/observed values which are close to unity.  Several of the model-predicted strengths agree with the observed values to within a few percent. 

Important exceptions to the above are lines associated with high ionization stages, such as N~V $\lambda$1240 and C~IV $\lambda$1549, where the predicted line strengths are nearly always significantly less than observed values. Attempts to match these lines by employing non-constant density distributions and SEDs other than blackbodies as well as moving the inner radius in by two orders of magnitude failed to solve the problem. It is possible that contamination from stellar emission is partially responsible for the unexpected observed strengths of these lines. For example, P-Cygni profiles are present in the spectra of IC~3568, NGC~5315 and NGC~5882 and likely account for a significant portion of emission in the N~V and C~IV lines (see Tables 3 and 4 of Paper~I). In addition, the slit passed through the central stars of IC~2165 and PB6, suggesting that stellar emission may be responsible for the excessive emission in at least those two cases. 

In addition to the N~V and C~IV line strengths, those of [S~II] were difficult to match in many instances. Except in the case of PB6, {where the model predicts a very different value of [S~II]$\lambda$6716/$\lambda$6731,} model predictions of $\lambda$6716 and $\lambda$6731 were always significantly less than their observed counterparts. By increasing the input S abundance in a model, the strengths of these lines could be raised but at the expense of elevating the strength of [S~III] $\lambda$9532 to an unacceptable level. Another way to force the agreement between observations and predictions for the [S~II] lines is to extend the computations farther out in radial distance from the central star. As mentioned above, all of our models were truncated at the radial distance where the observed and model [O~II] $\lambda$3727/H$\beta$ strengths agreed. However, because the ionization potential of neutral S is less than one Rydberg, [S~II] production can extend beyond the model cutoff and out into the neutral  zone. Several trials confirmed, however, that by extending the model outward both the [O~II] and [S~II] line strengths increased, forcing significant disagreement in the case of [O~II]. 

We also considered the possibility that shocked gas within the nebula was responsible for enhancing the strength of [S~II] above the value predicted by CLOUDY models. It is well known that in supernova remnants, for example, the spectra often show prominent emission from low ionization species such as [S~II] and [O~I], a result attributable to shock heating [cf. \citet{blair81,shull79,dopita77}]. Shock models computed by \citet{shull79} predicted [S~II]/[S~III] ratios of one to two orders of magnitude, depending upon the assumed value for the shock velocity. On the other hand, the value of this same ratio in a photoionized region such as the Orion Nebula is nearly the reciprocal of levels associated with shock heating \citep{esteban04}.

Clearly then, if shocks exist among our PN sample objects, they could explain the consistently high [S~II] emission. To pursue this idea we referred to the diagnostic diagrams published by \citet[Fig.~4]{frew10}, in which they plot log(H$\alpha$/[N~II]) versus log(H$\alpha$/[S~II]) for numerous PN, supernova remnants and H~II regions. These three object types clearly separate into their characteristic regions of the diagram. By placing the six PN for which we measured [S~II] emission onto the diagram we found that all six fell well within the PN domain and were clearly outside both the supernova remmant and H~II region domains. Thus, we feel that we can safely eliminate the possibility that some component of the observed [S~II] emission in our objects arises from shock heating.

It is probable, then, that the underproduction of [S~II] emission in the models is related to atomic data or processes currently employed by CLOUDY.  A likely candidate here is the dielectric recombination (DR) rate coefficient relevant to the S$^{+2}$$\rightarrow$ S$^+$ recombination. In a new paper by \citet{badnell15}, the authors employ a bootstrap method for deriving a revised DR rate of $3 \times 10^{-12}$cm$^3$ s$^{-1}$. This is roughly 100 times the rate derived by \citet{aldrovandi73}. Using this new rate they find that the fraction of S$^+$ in the gas increases at the expense of S$^{+2}$ and the ratio of [S~III]/[S~II] decreases by roughly a factor of 2. This is a promising result and could help explain why all of our models under predict [S~II] emission, since they did not include the new rate. However, Badnell et al. also point out that the increase in the DR rate does not entirely solve the problem of systematically lower-than-expected sulfur abundances observed in PN \citep{henry12}

As a quantitative measure of each model's success in matching its target set of observed spectral lines, we calculated an RMS value based upon the difference between observed and predicted line strengths,
$$RMS_{mod} = \left[\frac{1}{N}\sum_{lines} \left\{1-\left(\frac{model}{observed}\right)\right\}^2\right]^{1/2},$$
where the sum is over the line set containing $N$ lines in Tables~\ref{lsa} and \ref{lsb}; values for model/observed are given in the same tables for each object. $RMS_{mod}$ values should be compared to RMS$_{obs}$, which is related to the relative uncertainties of the observed line strengths:
$$RMS_{obs} = \left[\frac{1}{N}\sum_{lines} \left(\frac{\delta I}{I}\right)^2\right]^{1/2}.$$
In this second relation, $I$ and $\delta I$ are the reduced line intensity and its uncertainty, respectively, from Paper~I. The RMS values are listed in the last line of the {\it Observed} and {\it Model/Observed} columns in these two tables. Because of the particular problems with specific lines described above, we did not include C~IV $\lambda$1549, N~IV] $\lambda$1486 N~V $\lambda$1240 and any sulfur lines in the RMS calculations. A comparison of the two RMS values for each object clearly shows the offsets between models and observations are reasonably consistent with the line strength uncertainties, lending credibility to the models and their implications regarding central star properties. {However, we emphasize that despite the close agreement in terms of rms values, the fact that we failed to reproduce the strengths of those lines mentioned above that were excluded from the rms calculations diminishes any claim to a completely satisfactory model.}

Tables~\ref{abuna} and \ref{abunb} provide a comparison between the abundance ratios reported in section~3 of Paper~I with those required in the models to match the observed line strengths. Abundances in Paper~I were computed either by directly summing all of the observed ion abundances of an element or by applying an ionization correction factor to the sum of certain observed ion abundances (when not all occupied ionization levels can be observed directly). Except where noted, we report only the former values in Tables~\ref{abuna} and \ref{abunb}. The reader is referred to Paper~I for a detailed discussion of the observed abundances in our objects. {During the modeling process, we often adjusted abundances away from their starting (Paper~I) values in an effort to alter the cooling rate and thereby match important diagnostics such as the He~II $\lambda$4686/He~I $\lambda$5876 and [O~III] $\lambda$4363/$\lambda$5007 ratios, the nebular H$\beta$ luminosity and the strengths of the strongest emission lines, i.e., the most effective coolants. Consequently, we consider the {\it observed} abundances as best representing the true nebular composition.}



Tables~\ref{parama} and \ref{paramb} list the observed and model values for four important stellar properties and eight nebular properties.  The model stellar temperatures and bolometric luminosities represent values from our best model of each object. Except for PB6, the observed values for stellar temperature and bolometric luminosity are averages of the results of \citet{shaw85,shaw89,zhang93}, and \citet{frew08}, while the uncertainties were determined from the range among these published values, which we display in Table~\ref{paramc}. Each of these authors used the Zanstra method to estimate stellar temperatures. Shaw \& Kaler derived luminosities from reddening-corrected V band measurements, a temperature sensitive bolometric correction and statistical distances. Zhang \& Kwok found central star luminosities by first determining a star's mass from model tracks of known mass in a plot of 5~GHz brightness temperature of the nebula versus the stellar effective temperature, two observable quantities. Then model tracks of known mass in an H-R diagram plus knowledge of the star's effective temperature led to the luminosity. Frew derived luminosities using reddening-corrected absolute V magnitudes and bolometric corrections. Observed temperature and luminosity values for PB6 are from \citet{keller14}, who determined these parameters by modeling the stellar spectrum. The uncertainties quoted for PB6 represent our own estimates.  The stellar radii reported in the third row of Tables~\ref{parama} and \ref{paramb} are values in solar units based upon T$_{eff}$, log(L/L$_{\odot}$) and the Stefan-Boltzmann law. The stellar masses, again in solar units, are inferred by plotting each star on a theoretical HR diagram containing evolutionary tracks of post AGB model stars, as described in \S3. The observed masses were derived using the observed stellar temperatures and luminosities in Tables~\ref{parama} and \ref{paramb}, while the model masses result from plotting our model values for these parameters on the HR diagram. 

The lower section of Tables~\ref{parama} and \ref{paramb} show values related to the nebula itself. Observed H$\beta$ luminosities are derived from fluxes measured by \citet{cahn92} and distances reported in \citet{cahn92,zhang93} and \citet{kwitter06}. {Observed electron densities are taken from \citet{dufour15} and \citet[PB6 only]{garcia09}, while the predicted C~III], [S~II] and H$\beta$ densities were extracted from the model output. The total hydrogen density, N$_H$, served as the density input parameter for the model. Finally, observed and model nebular radii are taken from \citet{cahn92} and the model output, respectively.}

Here the agreement between observed and model results are generally satisfactory but with some discrepancies which are likely caused in part by distance uncertainties which impact observed values for stellar and nebular luminosities as well as nebular radii. We see excellent agreement for the temperatures, while the luminosities demonstrate significant disagreements in the cases of IC~2165 and NGC~5315. As we will describe below for both of these objects, numerous attempts to resolve these discrepancies were unsuccessful.

\subsubsection{Object by Object Analysis}

We now consider each of the eight objects individually and make some comparisons with earlier published work on each object. Statements regarding model quality ignore the frequent discrepancies involving the N~V, N~IV, C~IV and sulfur lines for reasons given earlier. We remind the reader that the observed temperatures and luminosities listed in Tables~\ref{parama} and \ref{paramb} are averages of individual measurements detailed in Table~\ref{paramc}.

{\it IC~2165}: Data in Table~\ref{lsa} show that all observed lines are reproduced by the model to well within 10\% of observed values. Observed and model abundances in Table~\ref{abuna} also compare well. Tables~\ref{parama} and \ref{paramc} show that the model T$_{eff}$ agrees well with observed values, {although somewhat lower than the temperature found by \citet{bohigas13} in their photoionization model of IC~2165}. At the same time, the model stellar luminosity is roughly one-fifth the observed value {in Table~\ref{parama} and well below the stellar luminosity value inferred by \citet{bohigas13}.} The observed temperatures and luminosities that we list in Tables~\ref{parama} and \ref{paramc} are consistent with earlier measurements reported by \citet{martin81}, who found a He~II Zanstra temperature of 102,000~K and a log(L/L$_{\odot}$) of 3.86. However, the nebular H$\beta$ luminosity computed by the models is very sensitive to the stellar luminosity, and our models could only match the former when a stellar luminosity well below the observed one was employed. Interestingly, \citet{gesicki07} computed a photoionization model of IC~2165 and found a stellar effective temperature of 155,000~K and a log luminosity relative to that of the Sun of 3.2, the former value being considerably higher than ours, but the latter value being in good agreement. Regarding density, the model density roughly matches the observed H$\beta$ density although not the C~III] density. {The Ar and N abundances in the model are roughly 1/3 and 1.3 times the observed value, respectively, while other model abundances are more closely aligned with their observed counterparts.} Finally, the model predictions for [O~III] temperature and nebular radius are consistent with their observed counterparts. 

{\it IC~3568}: Table~\ref{lsa} shows that the only discrepancy between model predictions and observations of emission lines involves the He~II $\lambda$1640 line, which we predict to be about 23\% stronger than the observed value. As indicated in Table~\ref{abuna}, all observed abundances but that of nitrogen agree well with the models, where the model required roughly three times more N to match the [N~II] lines. In addition, the observed and model parameters for IC~3568 in Table~\ref{parama} are consistent for both the central star and the nebula. {Our stellar temperature also closely matches the one employed by \citet{harrington83} in their photoionization model of IC~3568 based upon UV data from IUE plus optical data, although our luminosity is seven times higher than theirs.}

{\it NGC~2440}:  Observed and model line strengths agree reasonably well for this object, as seen in Table~\ref{lsa}.  Table~\ref{abuna} shows that the model abundances {also agree with the empirical determinations in Paper~I to within a factor of 2.} The notable exception is argon, where the model required only about one-third of the observed amount. In Table~\ref{parama} the model-predicted stellar temperature and nebular C~III] density, H$\beta$ luminosity, and [O~III] temperature all agree closely with observations, while the model value for stellar luminosity differs from observations by roughly a factor of 3 {and the model H$\beta$ density is significantly higher than the observed value.} There is sizable disagreement over the observed nebular radius, where \citet{cahn92} report 0.11~pc, while \citet{frew08} lists 0.18~pc. The radius is linearly dependent upon distance. We found three distance estimates for NGC~2440 in the literature, 2600~pc \citep{zhang93}, 1346~pc \citep{cahn92} and 1900~pc \citep{frew08} i.e., a range over a factor of 2, but clearly this uncertainty alone is not enough to explain the disagreement regarding nebular radii between our model and the observed values\footnote{NGC~2440 (and most other bipolars) have measured angular radii that depend on the exposure time used for the image. This is because the inner regions of the nebula are very bright compared to the extended structure. Moreover, owing to the complex ionization structures of bipolars, their sizes depend very strongly on the lines used for size measurements.}.

There are three previous photoionization models of NGC~2440 in the literature. The earliest one is by \citet{shields81} who computed a spherically symmetric constant density model using a Planck SED. They inferred a stellar temperature of 166,000~K and a nebular density of 4000~cm$^{-3}$. Later, \citet{bassgen95} fashioned a cylindrically symmetrical model and found a stellar temperature and luminosity of 125,000~K and 250~L$_{\odot}$, respectively. Finally, \citet{kh96}, like Shields et al., computed a model with spherical symmetry, a Planck SED having a T$_{eff}$ of 200,000~K and constant nebular density of 5700~cm$^{-3}$. Thus, all of these models, as well as the present one, are consistent with the notion that NGC~2440 is a high excitation nebula.  

{\it NGC~3242}: While there are no gross discrepancies between observed and model quantities for stellar and nebular parameters in Table~\ref{parama} for NGC~3242, we still had difficulty reproducing several of the observed line strengths (see Table~\ref{lsa}) and two of the nebular abundances (see Table~\ref{abuna}). For many of the lines in Table~\ref{lsa} the offset between observation and model for this PN is greater than 10\%. Since the slit for the observations of this object passed through regions of varying density, as indicated by changes in surface brightness, it is possible that the integrated light from the slit misrepresents conditions within the nebula. Regarding abundances, Ne and S show large disagreements. On the other hand, model line ratios which are indicative of stellar and nebular conditions, such as He~II $\lambda$4686/He~I $\lambda$5876 (excitation), [O~III] $\lambda$4363/$\lambda$5007 (gas temperature) etc., matched the observed values quite well. {\citet{henry00a} published a photoionization model of NGC~3242 in which they found $T_{eff}$=60,000K and log(L/L$_{\odot}$)=4.3, values significantly different from those used in our current model. However, the stellar parameter values presented here are more consistent with observed values then are those implied by their model.} Finally, we note that \citet{miller15} are analyzing spatially resolved segments along the slit for NGC~3242 in order to look for positional variations in properties such as gas density and chemical abundances.  

{\it NGC~5315}: Observed and model line strengths in Table~\ref{lsb} agree well for this object. However, several of the model abundances in Table~\ref{abunb} are considerably different from their observed counterparts; the model N and C abundances are greater than and less than the observed values from Paper~I, respectively. With regard to the stellar and nebular parameters for NGC~5315 in Table~\ref{paramb}, we found a somewhat higher value for T$_{eff}$ but a luminosity roughly one-third of the observed value. This model appears to be the first photoionization model published for NGC~5315.

{\it NGC~5882}: Our model for NGC~5882 reproduces most of the observed line strengths, as seen in Table~\ref{lsb}. The notable exception is He~II $\lambda$1640. This object's central star is known to exhibit a P-Cygni profile \citep{mendez88}, perhaps the cause of the additional observed emission in this line which is not duplicated in the nebular model. Abundance results listed in Table~\ref{abunb} show that observed and modeled abundances are mostly in agreement. An important exception is nitrogen, where our model's N abundance is roughly twice the observed value. The latter was determined in Paper~I using the ICF method, since lines of ions higher than N$^+$ could not be measured.  \citet{hkb04} reported a value for N/H of $1.81 \times 10^{-4}$, similar to our model value in Table~\ref{abunb}. Finally, in Table~\ref{paramb} we have good agreement for T$_{eff}$, nebular H$\beta$ luminosity and T$_e$([O~III]), and reasonable agreement for the central star luminosity. To our knowledge, this model is the first published photoionization model of NGC~5882.

{\it NGC~7662}: For the most part model lines reproduce the observed line strengths closely, as shown in Table~\ref{lsb}. Discrepancies include [O~III] $\lambda$4363 and He~II $\lambda$1640. Generally, model and observed abundances in Table~\ref{abunb} agree well. Finally, all observed stellar and nebular parameters in Table~\ref{paramb} are adequately reproduced by the model. Earlier photoionization models of NGC~7662 include those by \citet{bohlin78}, \citet{harrington82} and \citet{pequignot80}. In each case a stellar temperature was found to be on the order of 100,000~K. \citet{bohlin78} also found a luminosity of log(L/L$_{\odot}$) of 3.41, while the two models by \citet{harrington82} used values of 3.58 and 3.78 for this parameter. These earlier results are consistent with our model.

{\it PB6}: The modeling of PB6 was based upon our UV line intensities reported in Paper~I along with optical intensities and chemical abundances from \citet{pena98}. The use of Pe\~{n}a et al.'s data was necessary, since our optical intensities were of insufficient quality to allow confident comparisons with model line strengths. Thus, in Table~\ref{lsb} the observed UV intensities are from Paper~I, while intensities of lines beginning with $\lambda$3727 and extending longward are taken from Pe\~{n}a et al. One can see in Table~\ref{lsb} that the match between model and observation is satisfactory; an exception concerns the He~II $\lambda$1640 line strength. (However, note the large value of RMS$_{obs}$, which indicates the high level of uncertainties in the observed line intensities.) The agreement between the abundances shown in Table~\ref{abunb} {is less than satisfactory, as significant discrepancies exist in the cases of carbon, neon and sulfur. The model C value is roughly one-fourth of the observed level reported in Paper~I, while the model Ne and S abundance are roughly one-third and one-fourth of the observed amounts.} In Table~\ref{paramb} we see that both the model and observed nebular parameters are well-matched. The observed values of stellar temperature and luminosity from \citet{keller14} agree well with our model results. Previous photoionization models of PB6 have been published by \citet{hkh96} and \citet{pena98}. For a stellar temperature, Henry et al. found a value of 150,000~K while Pe\~{n}a et al. inferred a central star temperature of 154,000~K. Both values agree nicely with our current model. 

\section{Central Star Masses\label{properties}}

We now extend our discussion to the determination of the masses of the central stars of the planetary nebulae (CSPN) and their progenitors studied in this project. This is done by placing each CSPN on a theoretical HR diagram using our model-determined stellar temperatures and luminosities listed in Tables~\ref{parama} and \ref{paramb}. To this diagram we add a set of post AGB evolutionary tracks representing a range in progenitor mass. By comparing the stellar positions relative to the positions of nearby tracks, one can then estimate both the remnant and birth mass of each central star.

\begin{figure}
   \includegraphics[width=6in,angle=0]{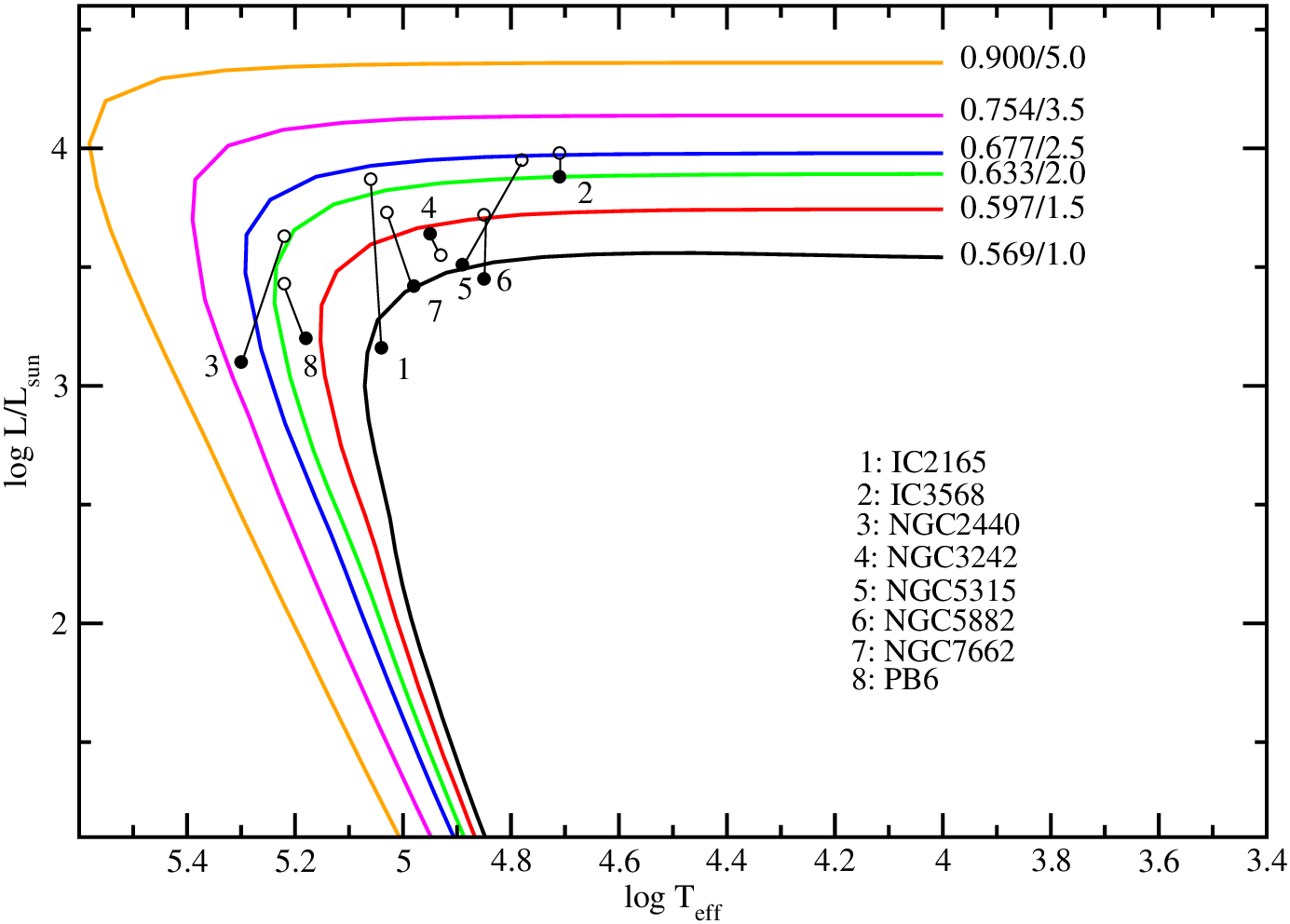} 
   \caption{Log L/L$_{\odot}$ versus log T$_{eff}$ for the eight central stars investigated here. Post-AGB model tracks are from \citet{vw94}, while the associated final/initial masses are shown at the right end of each track. Each filled symbol is identified by the name of the relevant PN, and its position is determined by the model-derived results of luminosity and temperature. Positions associated with {\it observed} luminosity and temperature values are shown with open symbols and are connected with thin lines to the model-derived objects. Uncertainties based upon the range of observed values of T and L  given in Table~\ref{paramc} are suppressed here to preserve clarity but are provided in Tables~\ref{parama} and \ref{paramb}.}
\label{cpn}
\end{figure}

The model-derived positions of our eight objects in the luminosity-temperature plane are shown with filled circles in Fig.~\ref{cpn}. We also show post-AGB model tracks from \citet{vw94} for their H-burning models with Z=0.016. Final/Initial masses related to each model are indicated at the low temperature end of each track and are based upon information related to each track provided by \citet{vw94}. The mass of each of our objects was determined by interpolating between bracketing model tracks; the resulting values are listed in the models column of Tables~\ref{parama} and \ref{paramb}. {We can see that there is large disagreement between observed and model initial masses for IC~2165, NGC~2440 and NGC~5315, moderately good agreement for IC~3568, NGC~5882 and NGC~7662, and good agreement for NGC~3242 and PB6. Finally, in a search of the papers mentioned in the previous section in which photoionization model results were reported for specific objects, we found that the only initial and final masses inferred directly from models pertained to IC~2165, for which both \citet{bohigas13} and \citet{gesicki07} found masses of roughly 2~M$_{\odot}$, in contrast to our value of 1~M$_{\odot}$.}  

For comparison we have also plotted the position (open circles) of each object according to its {\it observed} temperature and luminosity in Fig.~\ref{cpn}. The two symbols for each object are then connected by a thin solid line. Initial masses as implied by the observed temperature and luminosity were then determined in the same manner as described above for model-derived final masses, and are listed under the ``observed" column in Tables~\ref{parama} and \ref{paramb}. 

As a check on the effects of employing an alternate set of post AGB models on resulting masses, we constructed a second plot (not shown) similar to Fig.~\ref{cpn} but now containing tracks from the models of \citet{blocker95} and \citet{schonberner83} in place of those from \citet{vw94}. Differences in model-determined masses derived with this alternate set of tracks were on the order of 0.1~M$_{\odot}$ and therefore considered negligible.

How reliable are the mass values that we have determined? Estimating masses using the theoretical HR diagram and CSPN temperatures and luminosities, as we have done here, is vulnerable to large uncertainty due to the distance dependence of the observed luminosity. Observed luminosities are a function of the square of the distance to the PN, where PN distances can be poorly known. For example, an uncertainty of 50\% in distance translates to a factor of 2 (or 0.3 dex) uncertainty in luminosity and, according to the model tracks in Fig.~\ref{cpn}, a factor of 2 in mass. However, we derived our luminosities using photoionization models which were closely constrained primarily by distance-independent line ratios. Thus our final model-derived masses are likely to be much less affected by distance uncertainties than they would be if we simply used observed luminosities straight away. In addition to the problems created by distance uncertainties, we also recognize that our method for deriving progenitor masses is model dependent due to the use of post AGB tracks to estimate masses. This constitutes a systematic uncertainty and one that we can't quantify and correct for.

\begin{figure}
   \includegraphics[width=6in,angle=0]{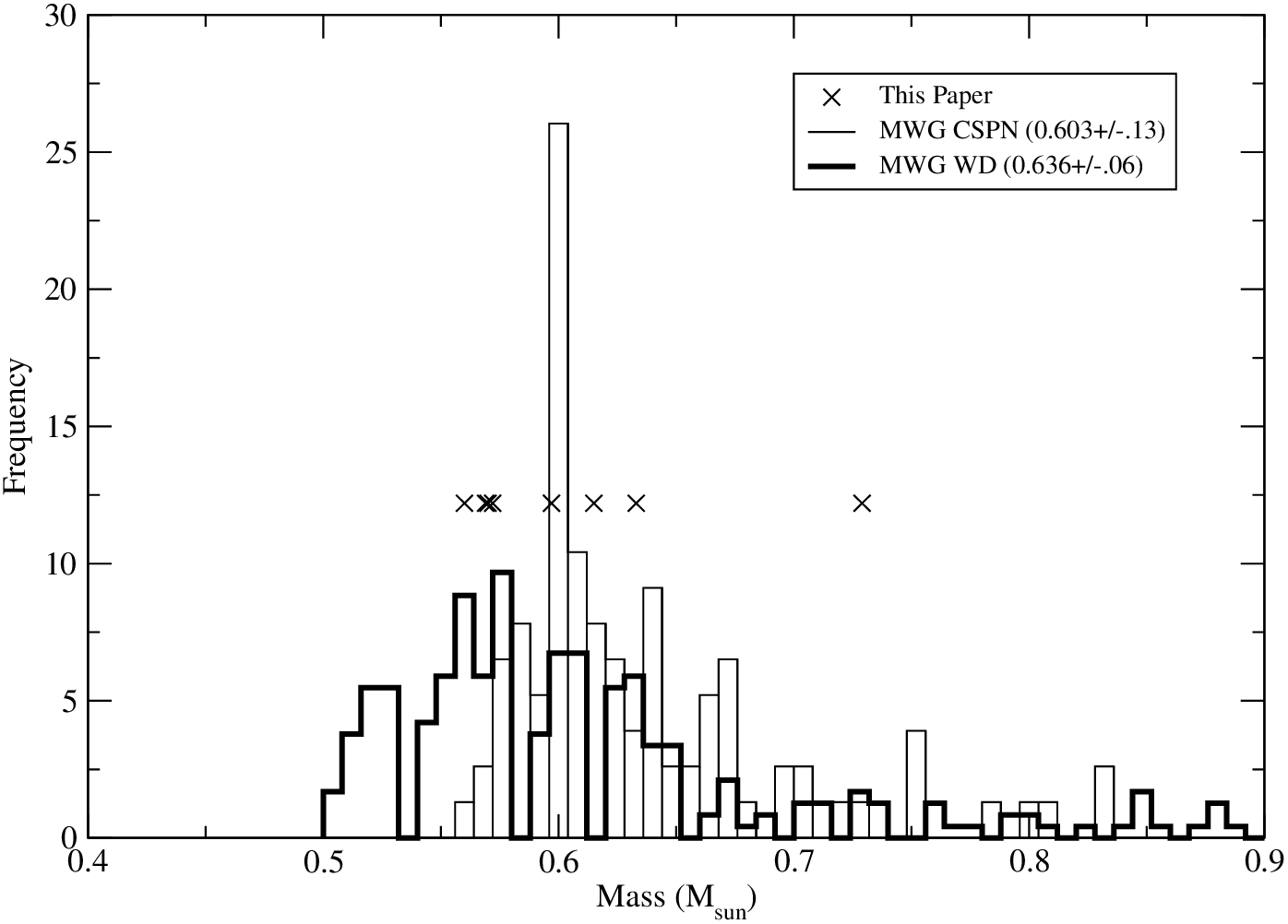} 
   \caption{Mass distribution of 297 white dwarf stars in the Milky Way disk from the Palomar Green survey presented by \citet[bold line]{liebert05},  and 91 Milky Way disk CSPN from \citet[faint line]{zhang93}. Averages and standard deviations in solar units are indicated in the legend for each group of objects. Ordinate values are normalized to the maximum in each survey. Horizontal positions of the central stars of our sample objects are {the model-determined masses from Tables~\ref{parama} and \ref{paramb} and} are shown with crosses (X).}
\label{hist}
\end{figure}

Figure ~\ref{hist} shows {our model-determined masses of the CSPN} sample (crosses) compared to the mass distribution for the sample of 297 white dwarf stars in the Milky Way disk from the Palomar Green survey published by \citet[bold line]{liebert05} and 91 CSPN in the Milky Way disk measured by \citet[thin line]{zhang93}. Mean values and standard deviations for these two samples are indicated in the legend. We see that the mass range of our objects agrees well with that of both samples in that seven of the eight CSPN are centered on the peak of the distributions of both the CSPN and white dwarf samples. The object with a remnant mass of 0.73 is NGC~2440. This comparison confirms the veracity of our derived masses at least in a statistical sense.

\section{Nucleosynthesis and Stellar Yields}

\subsection{Overview}

\begin{figure}
   \includegraphics[width=6in,angle=0]{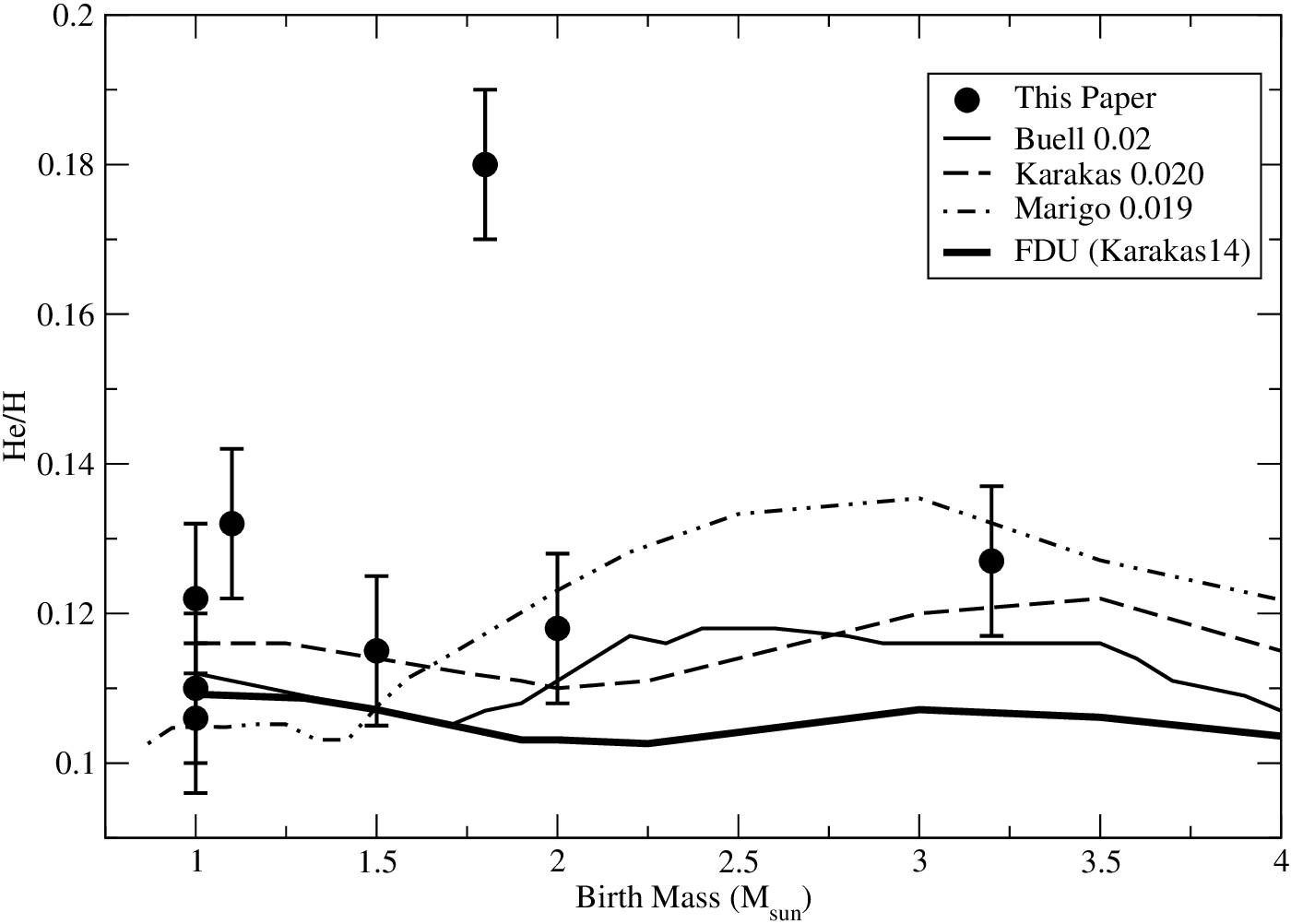} 
   \caption{He/H ratios versus stellar birth masses in solar units. {Observed He/H values for our sample} objects are shown with filled circles and error bars. The three thin lines show stellar evolution model predictions of PN abundance by \citet[thin solid line]{buell97}, \citet[dashed line]{karakas10} and \citet[dot-dashed line]{marigo01}, where model metallicities are indicated in the legend. The solid bold line shows the stellar model predictions of atmospheric He/H following first dredge-up from \citet[Table~1]{karakas14}.}
\label{he2h}
\end{figure}

\begin{figure}
   \includegraphics[width=6in,angle=0]{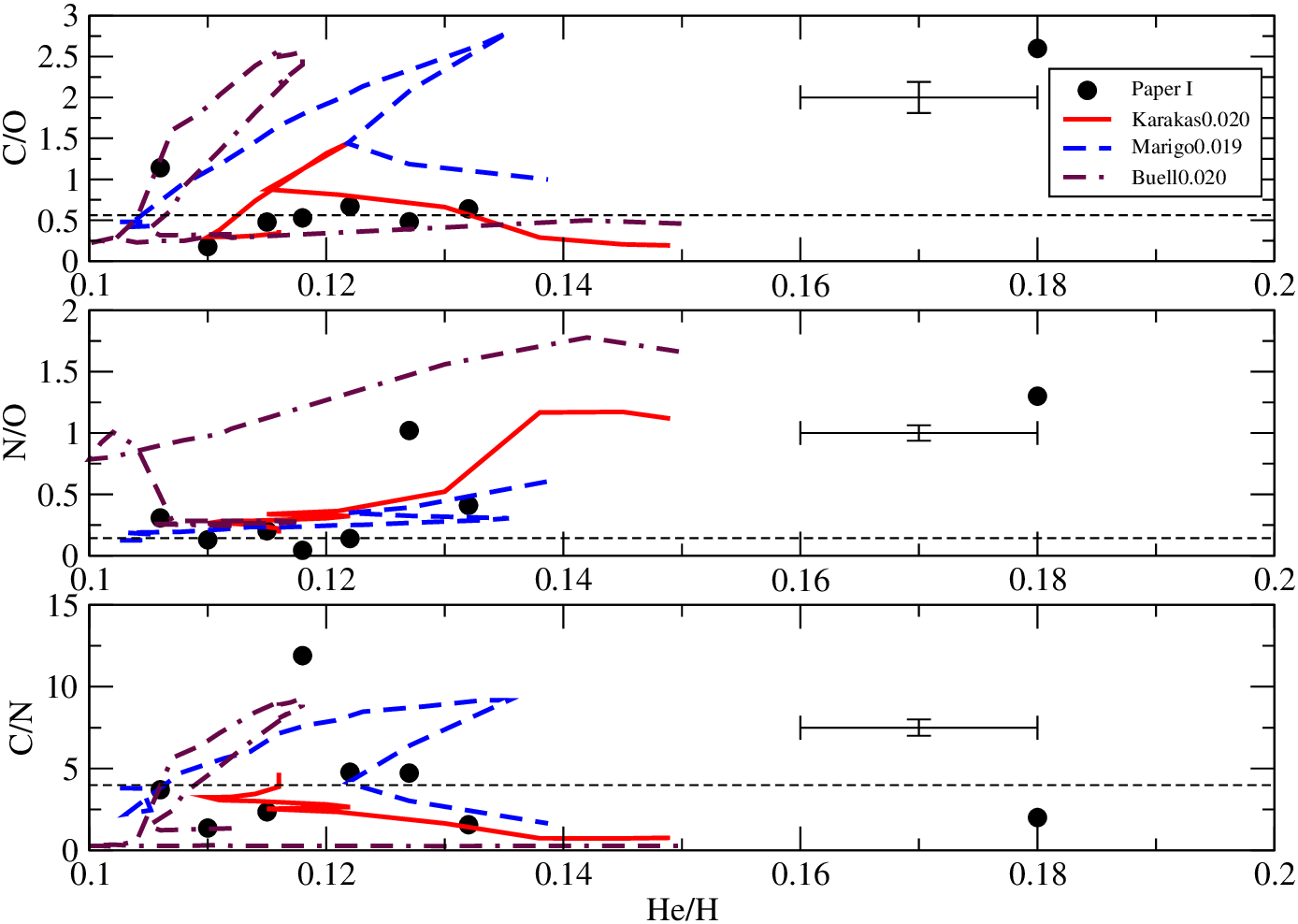} 
\caption{{\it Top panel}: C/O versus He/H. {Observed values for our sample} objects are indicated with filled circles and represent abundances reported in Paper~I, while model tracks from \citet{karakas10}, \citet{marigo01} and \citet{buell97} are designated by colors as indicated in the legend. For the model tracks, initial stellar mass increases generally from left to right. {\it Middle panel}: Same as top panel but for N/O versus He/H. {\it Bottom Panel}: Same as top panel but for C/N versus He/H. A sample error bar is shown in each panel. Solar values from \citet{asplund09} are indicated with dashed lines.}  
\label{cno1}
\end{figure}

\begin{figure}
   \includegraphics[width=7in,angle=0]{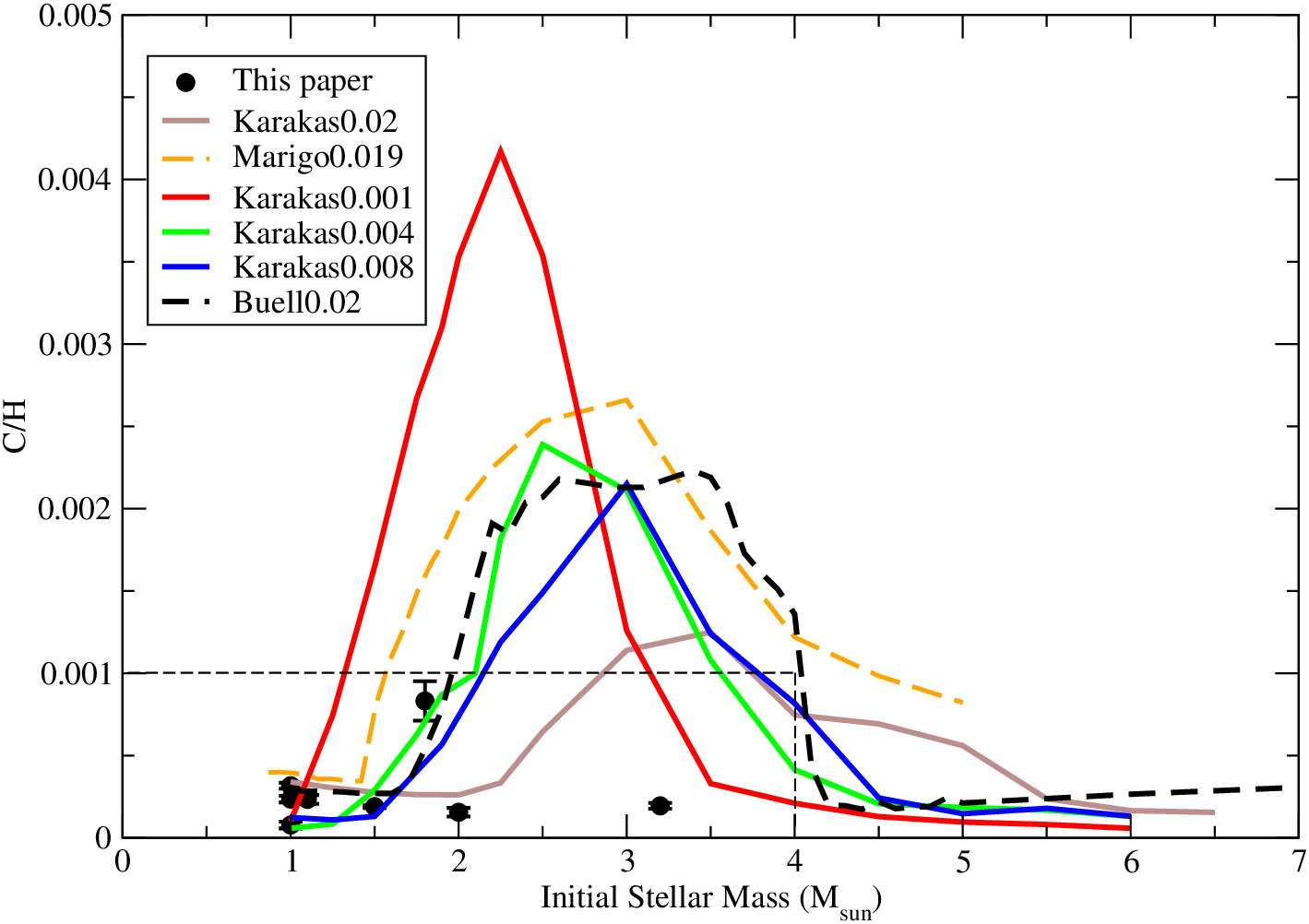} 
 \caption{Observed C/H number abundance ratio from Paper~I versus birth mass in solar masses of each PN progenitor star as derived in this paper (filled circles). Also shown are several tracks representing the predicted nebular abundances as a function of birth mass from theoretical models by \citet{karakas10}, \citet{marigo01}, and \citet{buell97}. The dashed lines show the region of the expanded plot in Fig.~\ref{c2hvstmass_enlarge}.}
\label{c2hvstmass}
\end{figure}

\begin{figure}
   \includegraphics[width=7in,angle=0]{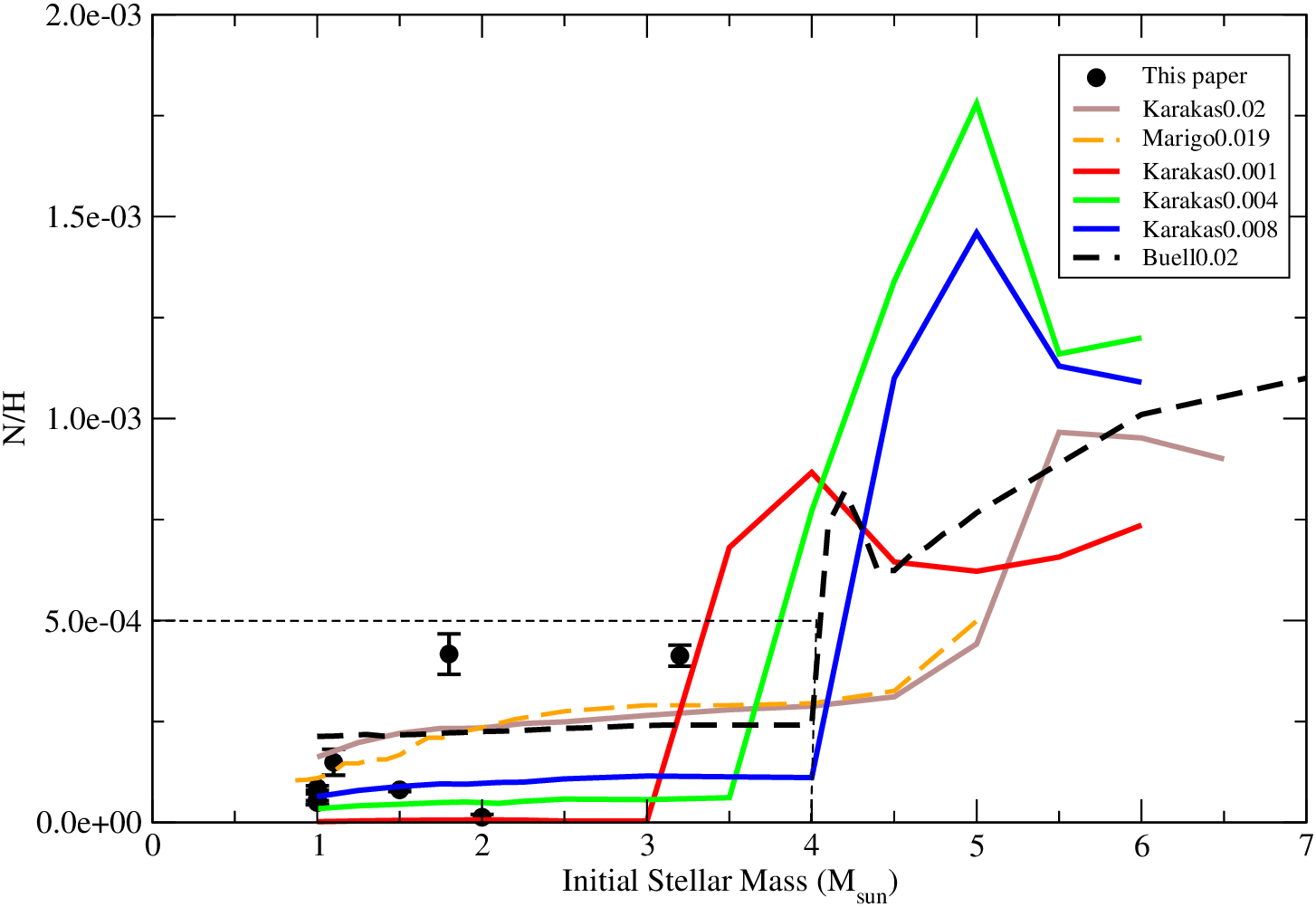} 
 \caption{Observed N/H number abundance ratio from Paper~I versus birth mass in solar masses of each PN progenitor star as derived in this paper (filled circles). Also shown are several tracks representing the predicted nebular abundances as a function of birth mass from theoretical models by \citet{karakas10}, \citet{marigo01}, and \citet{buell97}. The dashed lines show the region of the expanded plot in Fig.~\ref{n2hvstmass_enlarge}.}
 \label{n2hvstmass}
\end{figure}

\begin{figure}
   \includegraphics[width=7in,angle=0]{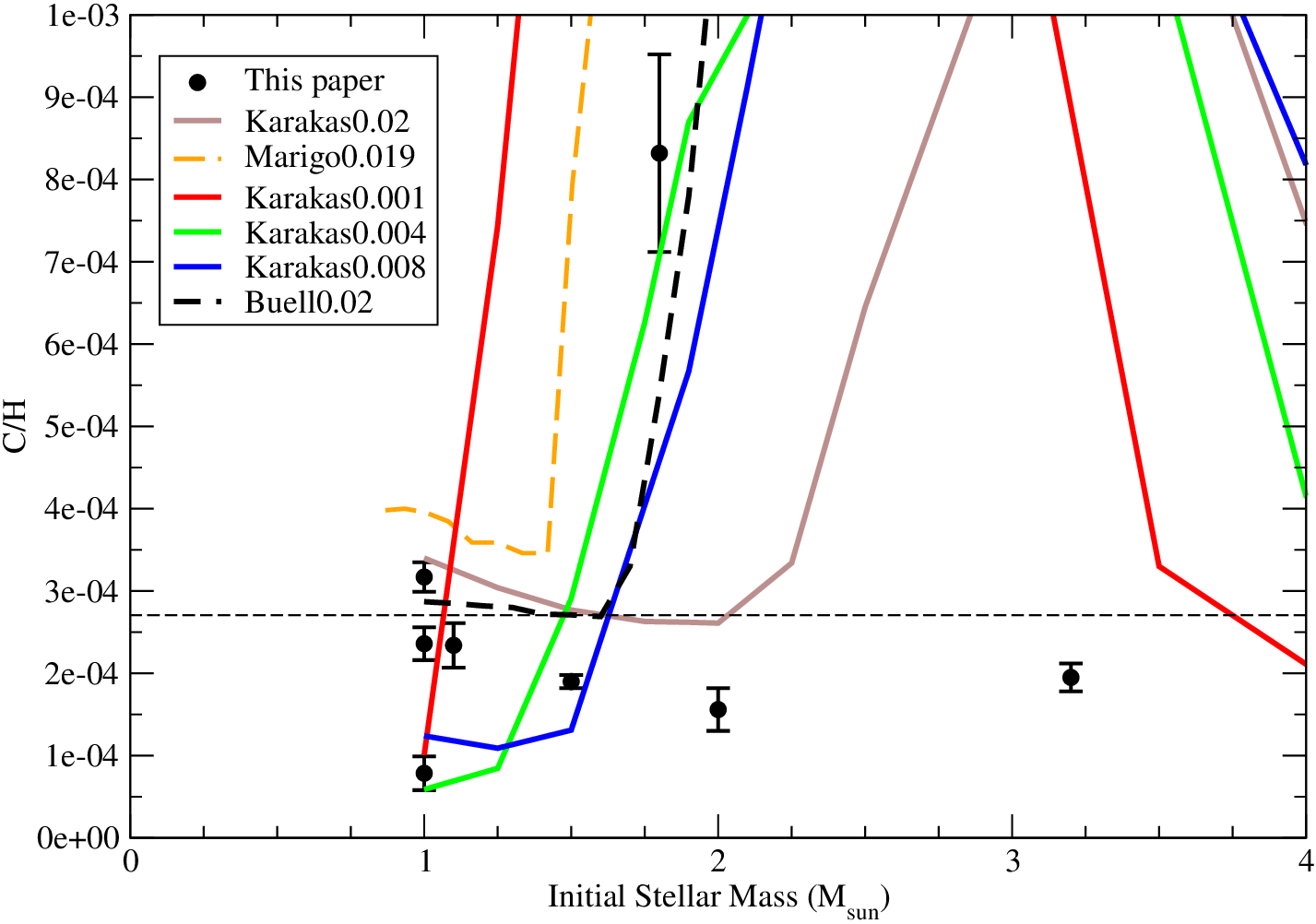} 
 \caption{Expansion of the region within the dashed lines of Fig.~\ref{c2hvstmass}, {where the C/H values are from Paper~I.} The solar value for C/H is $2.7\times10^{-4}$ \citep{asplund09} and is indicated with a horizontal dashed line.}
\label{c2hvstmass_enlarge}
\end{figure}

We now compare the nebular abundances and initial stellar masses of our sample objects derived here with the predictions of helium, carbon and nitrogen nucleosynthesis from published stellar evolution models. We acknowledge that this is a potentially hazardous endeavor, since both our inferred birth masses as well as published nucleosynthetic predictions are model dependent and constrained by observations containing inherent uncertainties. In addition, we are attempting to evaluate the stellar models based upon the study of only eight PN. Thus, our goal here is simply to provide a sense of the state of the art for inferring progenitor masses and predicting PN abundances by aligning the two sets of results and looking critically at the implications for being able to evaluate the relevance of the models and the usefulness of their stellar yields. {Note that in Figures \ref{he2h}-\ref{4plex2} discussed below, we chose to plot the {\it observed} abundance ratios from Paper~I. Because the model abundances often had to be adjusted during the modeling process to bring specific line strengths into agreement following the establishment of the major stellar and nebular parameters, we preferred to use the abundances from Paper~I for the following analysis. At the same time the stellar masses are those determined by our models unless otherwise noted.}

Results for relative helium abundance as a function of initial stellar mass are shown in Fig.~\ref{he2h}, where sample objects are indicated with filled circles and error bars. The solar He/H ratio is 0.085 \citep{asplund09}, so all eight PN appear to show helium enrichment with respect to that fiducial. [In fact large samples of PN show similar behavior \citep{kb94,hkb04}.] Also shown in Fig.~\ref{he2h} are stellar evolution model predictions of PN abundance by \citet[solid line]{buell97}, \citet[dashed line]{karakas10} and \citet[dot-dashed line]{marigo01}. {The basic properties of these model sets are briefly described in Appendix~A.} The prediction for atmospheric He/H after first dredge-up from \citet[Table~1]{karakas14} is indicated with the bold solid line. {Note that six of our PN (and their masses in solar units), IC~3568 (2.0), NGC~2440 (3.2), NGC~3242 (1.5), NGC~5315 (1.1), NGC~7662 (1.0) and PB6 (1.8) appear to have He/H ratios consistent with the occurrence of some 3rd dredge-up, while the He/H ratios of IC~2165 (1.0) and NGC~5882 (1.0) could be explained by 1st dredge-up only, although the final PN abundance predictions of Buell and Marigo are also consistent with the observed levels for these two objects.}

Note that for masses above 1.75~M$_{\odot}$ the PN abundance predictions exceed the prediction for first dredge-up, while below that threshold the two are similar. This is consistent with the idea that 3rd dredge-up produces further He enrichment in the PN with more massive progenitors {but is less effective} as the mass approaches a solar mass. Empirical support for this notion is provided by four of the five PN with initial masses less than or equal to 1.5~M$_{\odot}$; NGC~5315 at 1.1~M$_{\odot}$ is the exception. 

Fig.~\ref{cno1} contains three plots, each showing a comparison of observations and model predictions for three element ratios, C/O, N/O, and C/N, versus He/H. C/O and N/O are measures of C and N enrichment relative to progenitor metallicity, while C/N indicates the relative effectiveness of 3rd dredge-up versus hot bottom burning. Filled circles indicate positions of our eight objects based upon abundances determined in Paper~I. The comparison model tracks show predicted PN abundances and are taken from work by \citet{karakas10,marigo01} and \citet{buell97} for metallicities of 0.02, {0.019}, and 0.02, respectively. [While the value of 0.02 was formerly taken to represent the solar metallicity, more recently \citet{asplund09} claim that value to be 0.0134.] The dashed line in each plot indicates the solar value taken from Asplund et al. (2009) for the ratio on the vertical axis. Their solar value for He/H is 0.085, located slightly to the left of the plot's minimum. The extreme object with He/H=0.18 is PB6.

{In the case of C/O in the top panel, the Buell and Karakas tracks agree well with observations, while the Marigo model appears to predict too much nebular carbon.} An interesting result is the absence of C enrichment in all but two PN (IC~2165 and PB6). {This suggests that the observed helium enrichment noted above is from first dredge-up, where C enrichment is expected to be negligible.}  

In the middle panel observed N/O is compatible with the Marigo and Karakas models but is not reproduced in Buell's model. N/O is often observed to be positively correlated with He/H \citep{dufour91,dufour15}, especially for a sample of Type~I PN. However, such a trend is not apparent here, most likely because NGC~2440 and PB6 are the only two of our eight objects which are members of the Type~I Peimbert class. Finally, in the bottom panel each model seems to be in agreement with a few of the points but not the majority of them. Our conclusion based upon Fig.~\ref{cno1} is that while the models don't closely predict the nebular abundance ratios of C/O and N/O, they do at least span the space occupied by the observed points, thereby lending some support for the models. Indeed part of the difference between observation and theory may be related to the actual metallicity differences between the models and our sample objects.

Figures~\ref{c2hvstmass} and \ref{n2hvstmass} are plots of C/H and N/H versus stellar birth mass, respectively. Our sample objects are shown as filled black circles. In some cases, particularly in the full-scale plots, the formal error bars of the present abundance determinations are too short to extend beyond the symbol\footnote{Although our measurement errors may be small, our disagreements with credible, previously published abundances can be considerable. Some of these differences are explored in Paper~I.}.  The observed C/H and N/H values on the vertical axis are again taken from Paper~I, while the lines showing model tracks are based upon the same computations as the ones mentioned above by \citet{karakas10,marigo01}, and \citet{buell97} for the metallicities indicated in the legend.

All model tracks in Figure~\ref{c2hvstmass} show the marked increase in C/H between 1-4~M$_{\odot}$ resulting from 3rd dredge-up. The decline in this ratio beginning near 3~M$_{\odot}$ is due to hot bottom burning, where material at the base of the H-rich convective envelope reaches temperatures on the order of 10$^8$~K as it enters the H-burning shell \citep{lattanzio04,marigo03,herwig05}. As H is then burned to He via the CN cycle, C from 3rd dredge-up is converted to N as cycle equilibrium is reached. Thus with increasing mass the decline in C/H in Fig.~\ref{c2hvstmass} coincides with the rise in N/H in Fig.~\ref{n2hvstmass} beyond 3~M$_{\odot}$, as these products are mixed up into the atmosphere by convection.

In the following discussion for the first time we directly compare observed C and N abundance results with the model predictions of PN nebular abundance ratios by \citet{karakas10} specifically as functions of progenitor mass and metallicity. Karakas' models are chosen here because they currently appear to be the most recent and complete theoretical study of element production in terms of ranges in mass and metallicity; we are in no way disparaging the work of earlier investigators. We first compare the observed and model-predicted abundances of carbon. A similar analysis of nitrogen follows that discussion. 

\subsection{Carbon\label{carbon}}

\begin{figure}
   \includegraphics[width=6in,angle=0]{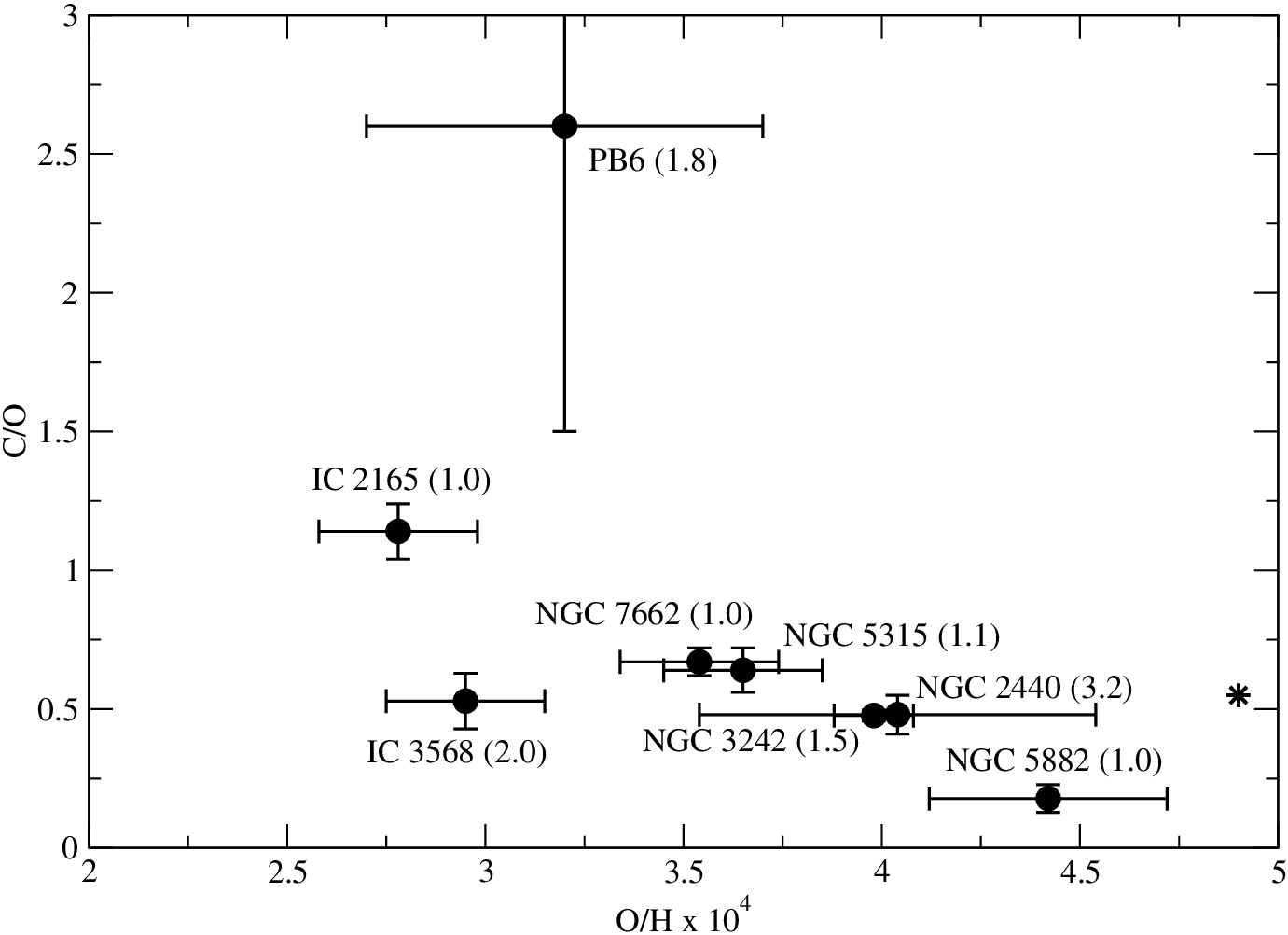} 
\caption{Observed C/O versus O/H abundance ratios from Paper~I for our eight sample objects. Each object is identified by its name and progenitor mass in solar masses from Table~\ref{abuna} or \ref{abunb} in parentheses. The solar values from \citet{asplund09} are indicated by an asterisk.}  
\label{c2ovo2h}
\end{figure}

\begin{figure}
\includegraphics[width=6in,angle=0]{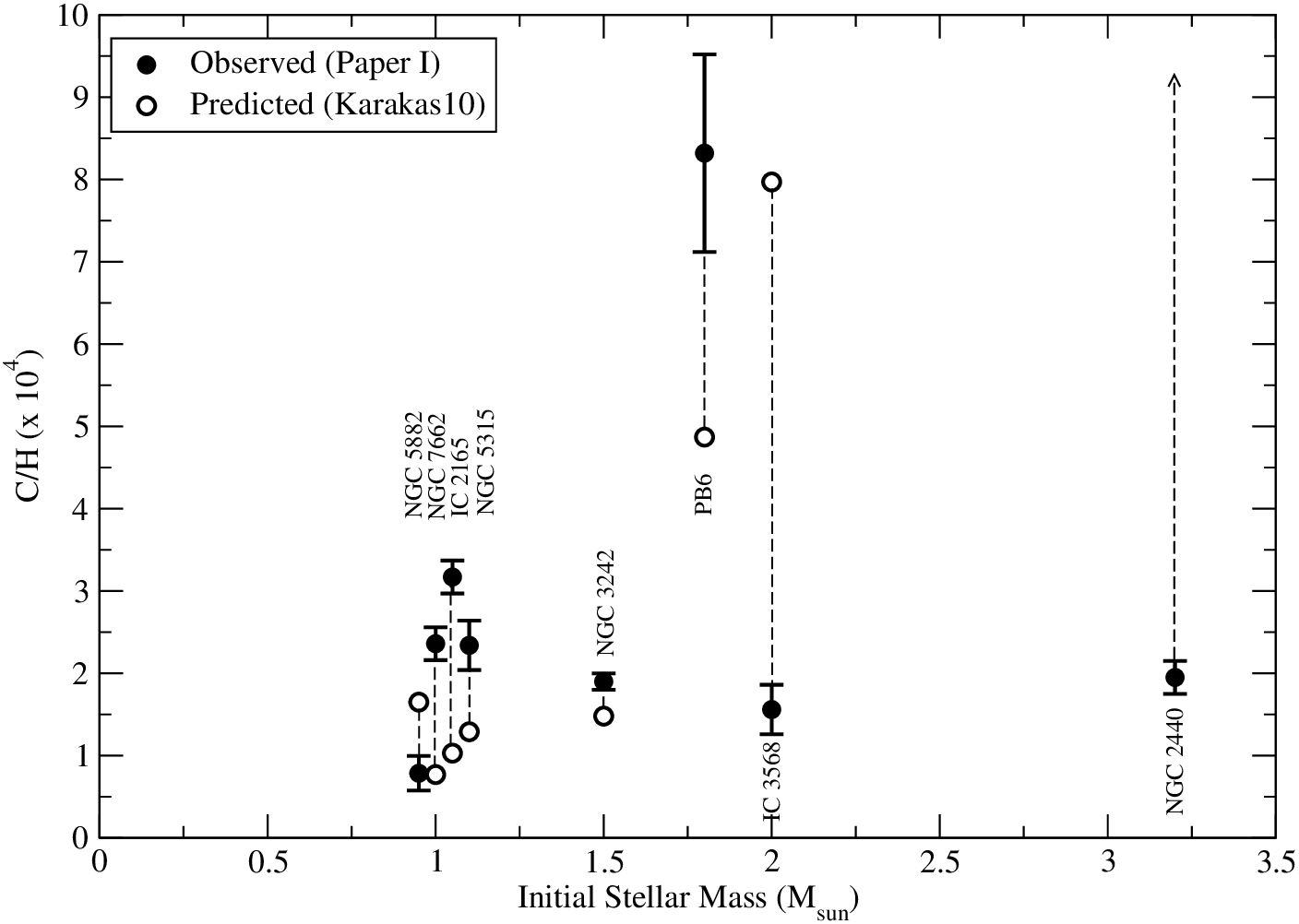}
\caption{C/H ratios versus initial stellar mass as derived here and reported as ``model" mass in Tables~\ref{parama} and \ref{paramb} in solar units. Observed C/H abundance ratios taken from Paper~I and model-predicted abundance ratios from \citet{karakas10} appear as filled and open circles, respectively. A vertical dashed line connects the two symbols of each object. Objects are identified by name. NGC~5882, NGC~7662 and IC~2165 have masses of 1~M$_{\odot}$ but are offset slightly from each other for clarity.}
\label{4plex1}
\end{figure}

\begin{figure}
   \includegraphics[width=7in,angle=0]{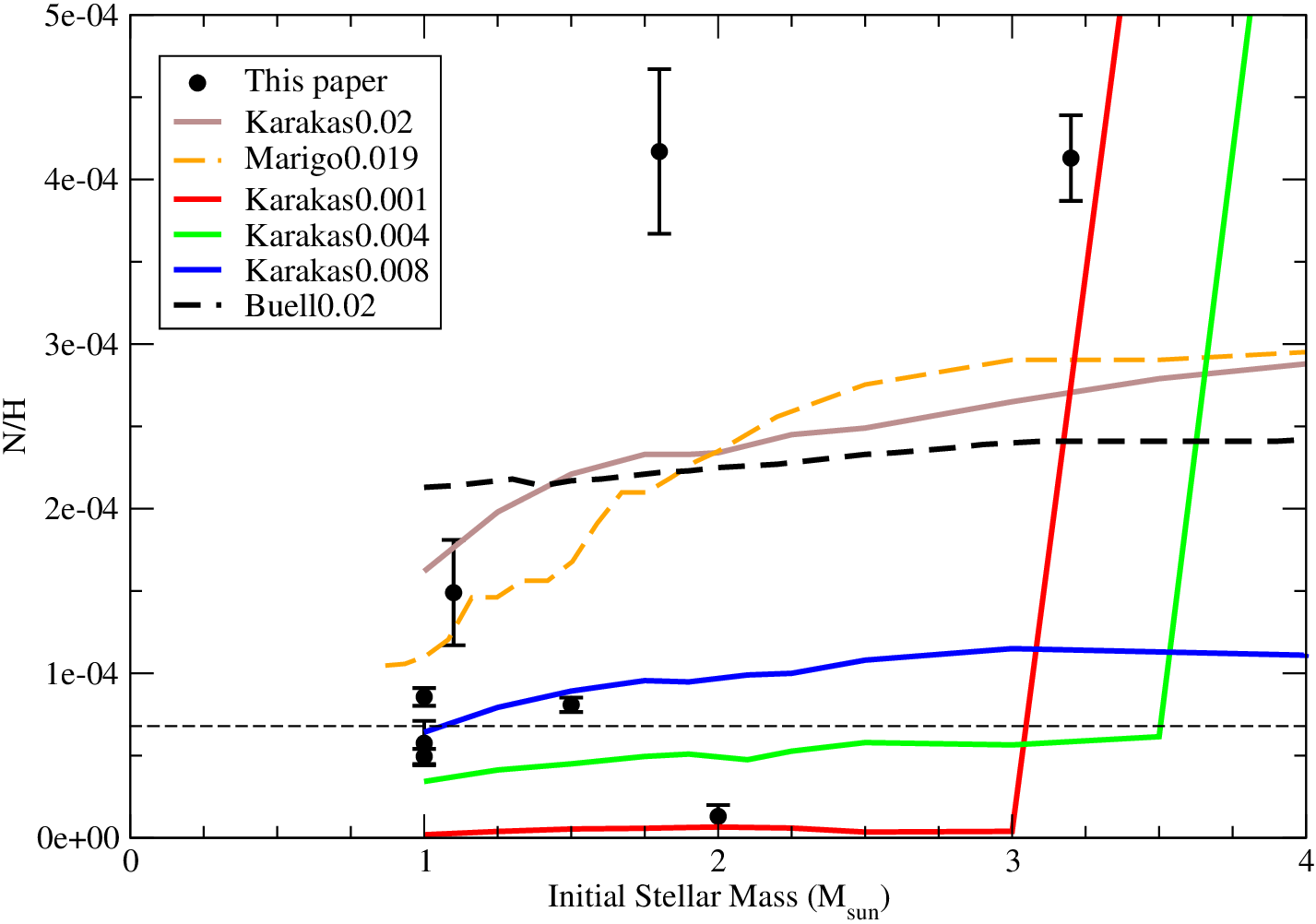} 
 \caption{Expansion of the region within the dashed lines of Fig.~\ref{n2hvstmass}, {where the N/H values are from Paper~I.} The solar value for N/H is $6.8\times10^{-5}$ \citep{asplund09} and is indicated with a horizontal dashed line.}
 \label{n2hvstmass_enlarge}
\end{figure}

With the exception of NGC~2440 at 3.2~M$_{\odot}$, the C/H values of our sample objects displayed in Fig.~\ref{c2hvstmass} are roughly spanned by model predictions relevant to low mass progenitors. The region in Fig.~\ref{c2hvstmass} within the two dashed lines is enlarged in Fig.~\ref{c2hvstmass_enlarge} and shows the vertical displacements more clearly.  Assuming a solar value of $2.7 \times 10^{-4}$ \citep{asplund09}, we see that C/H for this group ranges from $7.86 \times 10^{-5}$ or 0.29 solar for NGC~5882 to $8.32 \times 10^{-4}$ or 3.08 solar for PB6. Figure~\ref{c2ovo2h} is a plot of C/O versus O/H for our eight objects. Each object is identified by name, and the initial mass of the central star, as derived above, is given in parentheses. Excluding PB6, there is a clear negative trend such that C/O decreases as metallicity (gauged by O/H) increases. At lower metallicities C/O is likely to be higher simply because the oxygen abundance is already reduced relative to the carbon that is produced by the star. On the other hand, \citet{karakas02} have shown that for similar masses the 3rd dredge-up parameter, $\lambda$, is a decreasing function of stellar metallicity. Therefore, as metallicity is reduced one would expect greater amounts of C to be dredged up and mixed into the atmosphere, causing C/O to rise. The observed trend in our data shown in Fig.~\ref{c2ovo2h} supports both of these explanations. On the other hand, PN abundance predictions of C and O by \citet{karakas10} and \citet{marigo01} also suggest that at constant metallicity C/O is positively correlated with initial mass. Unfortunately, the metallicity range of our objects (0.57-0.90 times the solar value for O/H) is too large to test for this effect.

Next, we make a detailed comparison of our objects' carbon abundances with Karakas' PN abundance predictions. Karakas computed models for several masses between 1 and 6 solar masses and for three metallicities relevant to our study, i.e., 0.02, 0.008, and 0.004. Since our objects' masses and metallicities generally did not match the mass and metallicity values for any particular model, we performed interpolations in both parameters to estimate the model-predicted carbon abundance for a specific object.  The birth masses were the values computed earlier in this paper and listed in Tables~\ref{parama} and \ref{paramb}. In the case of metallicity,  we converted model metallicity values to values of O/H\footnote{Oxygen abundances in Karakas' models appear to be altered only slightly during a star's evolution, so we assumed that O/H represents a reliable gauge of metallicity. We considered using Ne/H as an alternative indicator of metallicity. However, model predictions suggest that neon abundance can be altered substantially during post-main sequence evolution.} by scaling the O/H solar abundance from \citet[Z=0.02]{anders89}\footnote{The same set of solar abundances used by Karakas.}

Observed and predicted values for C/H are compared in Table~\ref{comparisons}. Note that objects are arranged from top to bottom in order of increasing birth mass. For each object identified in the first column, we show the birth mass of the progenitor star {based upon our photoionization models} and the observed O/H used to interpolate between models within metallicity space. The next three columns show the observed and predicted abundances along with the ratio of the predicted to observed values. Results in columns 4 and 5 for each object are also shown graphically in Fig.~\ref{4plex1}, where we have plotted C/H versus progenitor mass. Filled and open circles represent observed and predicted values, respectively. Since IC~2165, NGC~5882, and NGC~7662 each has a mass of 1~M$_{\odot}$, the three are slightly offset from one another for clarity. Vertical dashed lines connect the observed and predicted symbols for each object. The predicted value for NGC~2440 is $17.1 \times 10^{-4}$ and falls outside the vertical limits of the plot.

We can see clearly {in column~6} that the largest offsets between observed and predicted carbon abundances involve IC~3568 and NGC~2440, two PN with inferred stellar birth masses equal to or exceeding about 2~M$_{\odot}$, when the C abundance is expected to be greatly affected by 3rd dredge-up. On the other hand, predicted abundances for the remaining six objects with progenitor star masses below this limit differ from observed values by at most a factor of 3. Intuitively, this behavior might be expected. Carbon abundances in PN of larger mass progenitors are likely to be more sensitive to the uncertainty of the dredge-up parameter as well as the number of dredge-up episodes in the models. {\it Thus, our results suggest that Karakas' models reasonably predict PN nebular abundances for progenitor masses below about 1.5~M$_{\odot}$, but over-predict C/H by substantial amounts in the mass range of LIMS in which 3rd dredge-up plays a significant role.} Clearly, this is only a preliminary conclusion, as more high mass LIMS need to be analyzed with similar methods to confirm this result.


\subsection{Nitrogen\label{nitrogen}}

\begin{figure}
   \includegraphics[width=6in,angle=0]{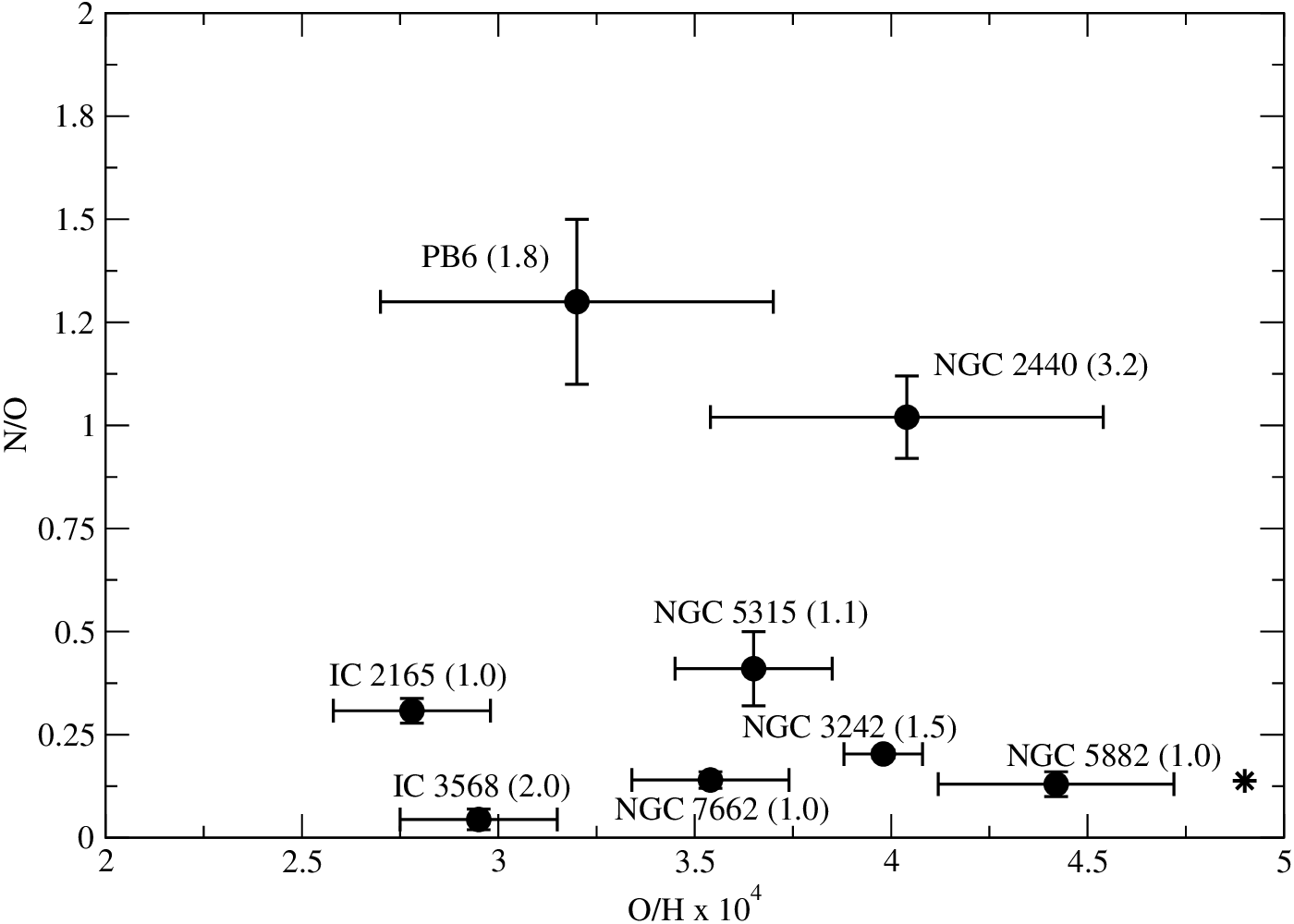} 
\caption{Observed N/O versus O/H abundance ratios from Paper~I for our eight sample objects. Each object is identified by its name and progenitor mass in solar masses from Table~\ref{abuna} or \ref{abunb} in parentheses. The solar values from \citet{asplund09} are indicated by an asterisk.}  
\label{n2ovo2h}
\end{figure}

\begin{figure}
\includegraphics[width=6in,angle=0]{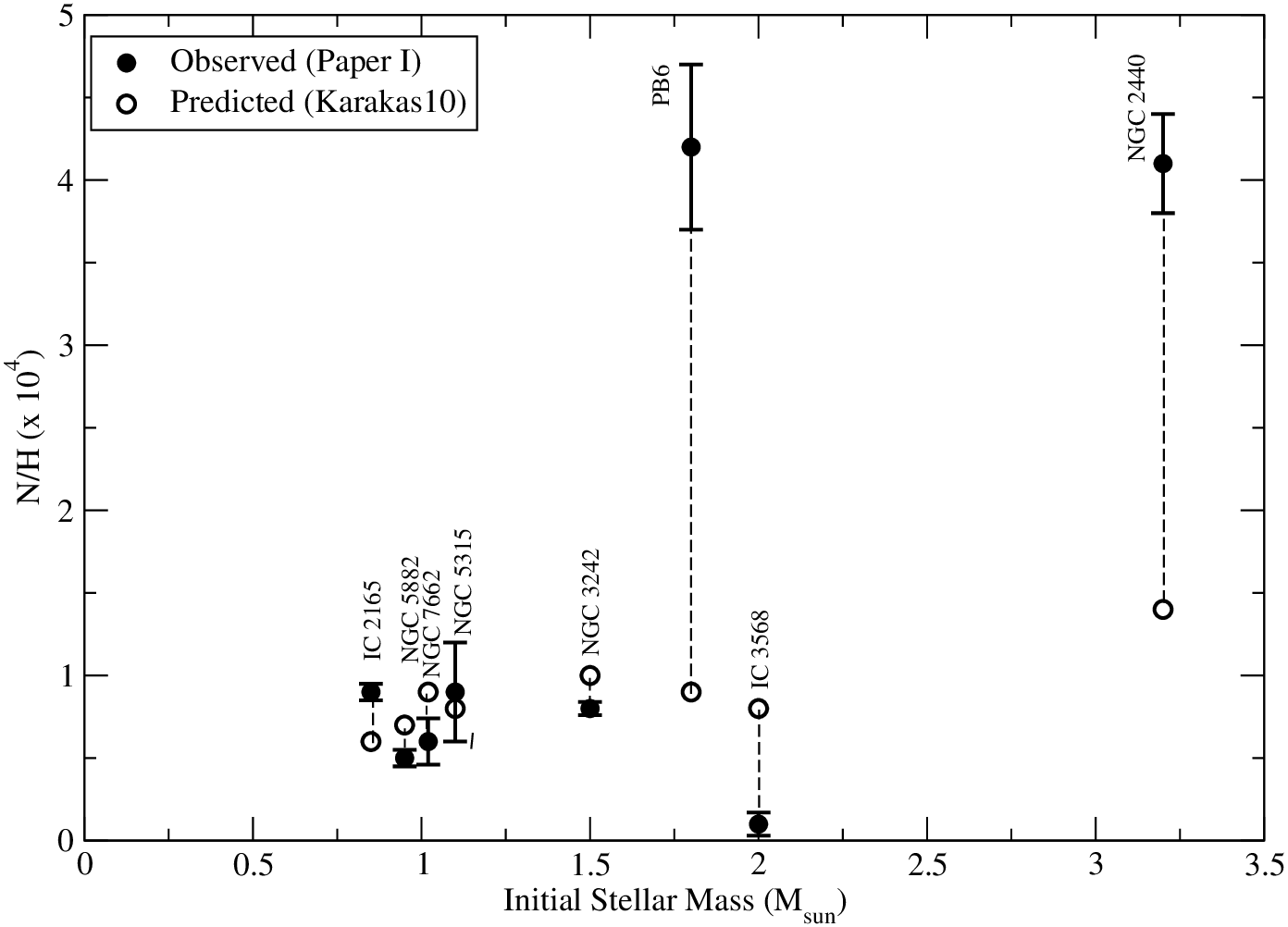}
\caption{N/H versus initial stellar mass as derived here and reported as ``model" mass in Tables~\ref{parama} and \ref{paramb} in solar units. Observed N/H abundance ratios taken from Paper~I and model-predicted abundance ratios from \citet{karakas10} appear as filled and open circles, respectively. A vertical dashed line connects the two symbols of each object. Objects are identified by name. NGC~5882, NGC~7662 and IC~2165 have masses of 1~M$_{\odot}$ but are offset slightly from each other for clarity.}
\label{4plex2}
\end{figure}

The analysis just described for carbon was repeated for nitrogen. The results appear in Table~\ref{comparisons} and Figs.~\ref{n2ovo2h} and \ref{4plex2} (again, the three objects of 1~M$_{\odot}$ are slightly offset from one another for clarity). In contrast to what we saw with carbon (see Fig.~\ref{c2ovo2h}), we now see no apparent mass-metallicity behavior regarding PN nitrogen abundances in Fig.~\ref{n2ovo2h}; and analogous to the case for carbon, there is no correlation between N/O and stellar mass. In Fig.~\ref{4plex2} for birth masses of $\le 1.5$, the differences between observed and predicted nebular abundances are rather small. Above this level, for PB6 and NGC~2440 the predicted abundances are far below the levels observed. {(Recall that these two objects are classified as Peimbert Type~I PN.)}

The generally small abundance predictions in Fig.~\ref{4plex2} are likely the result of the progenitor masses of our objects being below the mass range where large amounts of N are produced through hot bottom burning. We see this clearly in Fig.~\ref{n2hvstmass}, where the predicted rise in N begins around 3~M$_{\odot}$ for a metallicity of 0.001 and at greater masses for higher metallicities. For masses below this limit, N is produced primarily during first dredge-up, resulting in smaller amounts of enrichment; the levels here are evidently the value at star formation plus the dredge-up contribution. Model uncertainties in these cases are less likely to have an effect on the predicted N enhancements compared with the situation in which hot bottom burning operates.  

{\it We conclude that model predictions of N enrichment through dredge-up at masses below 3~M$_{\odot}$ in the Karakas models are consistent with observations.} Investigations of progenitor stars of He/H- and N/O-rich PN of Type-1 Peimbert class (such as NGC~2440) are now in order, since those objects are believed to have masses exceeding the lower limit for hot bottom burning. Such studies will allow us to evaluate the model predictions for stars that produce significant amounts of nitrogen during their AGB stage.

\section{Summary and Conclusions}
  
This paper follows up on the study published by \citet{dufour15} of ten bright planetary nebulae. The novelty of their work was the use of co-spatial spectrophotometric measurements gained through the use of HST/STIS and spanning the wavelength range of 1150--10270\AA~to infer abundances of many elements, but especially those of carbon and nitrogen. 

The goal of the present work is to compare the observed He, C and N nebular abundances with predictions of the same made by published AGB and post-AGB stellar evolution models available in the literature, where the same models simultaneously predict stellar yields of these two elements. We recognize that agreement between observed and model nebular abundances is necessary but not sufficient to confirm the correctness of the yield predictions themselves. Yet it is one important check on the models. Testing of the predicted yields in chemical evolution models can then further confirm their applicability.

In the past comparing nebular abundances with model predictions has been done by plotting observed PN abundances along with model tracks representing the predictions in an element versus element graph {[cf. \citet[Figs.~5.4 and 5.5]{buell97}, \citet[Fig.~4]{henry00a}, \citet[Fig. 11]{marigo01}].} Here for the first time we have carried out the analysis by coupling the abundances from each PN with its progenitor star mass. 

To obtain these masses, we have used the measured line strengths from Paper~I as primary constraints to compute a detailed photoionization model for eight of their objects: IC~2165, IC~3568, NGC~2440, NGC~3242, NGC~5315, NGC~5882, NGC~7662 and PB6. Because these models require a stellar temperature and luminosity as input, successfully matching the observed line strengths with the model's output line strengths helps us to confirm these stellar properties.\footnote{{The level of our success in matching the observed line strengths from Paper~I was evaluated} by computing and then comparing observed and model RMS values representing most of the important lines in the spectrum (see \S2.1).} In most cases these spectroscopically determined properties compared favorably to previously published but more distance-dependent measurements available in the literature. {Employing our model-determined central star temperature and luminosity from Tables~\ref{parama} and \ref{paramb}}, we then plotted the eight objects on a theoretical HR diagram containing post-AGB model tracks, where each track is associated with a specific central star and progenitor star mass. The mass for each of our objects was established by carefully interpolating between model tracks.

With progenitor masses and C and N abundances {from Paper~I} in hand, we proceeded to make a detailed comparison between model-predicted and observed values for each PN. The former were determined from the models by \citet{karakas10}, using our inferred progenitor mass and observed nebular metallicity to chose the best model from which to take the prediction. Often this meant interpolating between masses and/or metallicities of specific models in order to match the conditions of an object. 

The salient points that emerge from this exercise are:

\begin{enumerate}
\item Matching observed line strengths with photoionization models provides a good method for determining PN central star temperatures and luminosities, and by extension, central star and progenitor star masses. This is not a new result, of course, but our experiment here with eight objects appears to show that this method can nearly always be used to infer reliable values for these stellar parameters that are relatively distance independent.

\item The central star mass distribution of our objects is very consistent with the distribution of white dwarf and central star masses from other studies.

\item Most of our objects show little if any C enrichment, indicating that the effects of 3rd dredge-up are negligible, as expected for our range of stellar masses.

\item A plot of C/O versus O/H abundance ratios for our objects indicates the presence of an inverse correlation.This may be due either to a metallicity sensitivity of the third dredge-up parameter predicted by model results for AGB stars or simply to the reduction of oxygen at lower metallicities.

\item For progenitor masses below about 1.5~M$_{\odot}$, stellar model predictions of nebular C enrichment appear to be consistent with the observations. However, the observed abundances of C in the few objects above this threshold seem to indicate that the stellar models overpredict nebular C abundances. This discrepancy could be explained by the values of the dredge-up parameter and/or the number of pulses used in the models.

\item Model predictions for N enrichment in the PN are consistent with observed levels for progenitor masses below 3~M$_{\odot}$. Hot bottom burning becomes more effective above this threshold, so concentrating future work on PN with progenitor masses above this level are necessary to check further on the model predictions.

\item {PB6 stands apart from the other seven PN in terms of its abundance pattern. Its unusually large He/H, C/O and N/O ratios clearly point to a history of enrichment by the progenitor star. This, coupled with the presence of a [WC] central star \citep{garcia09}, makes this an interesting object in terms of understanding the nucleosynthesis that occurred during its post main sequence lifetime. Figures \ref{he2h}, \ref{4plex1} and \ref{4plex2} also demonstrate that current post AGB models fail by a large margin to predict the levels of He, C and N currently observed for this object.} 
 
\item As nearly as we can tell, this work presents the first published photoionization models of NGC~5315 and NGC~5882.

\end{enumerate}

Note that points {5, 6 and 7} above apply only to the models by \citet{karakas10}, since we did not test other published model sets. We emphasize the importance of extending the study of carbon and nitrogen production by LIMS beyond the central star progenitor mass of 1.5~M$_{\odot}$ by emphasizing that, using a \citet{salpeter55} initial mass function, LIMS greater than this mass represent roughly 54\% of all stars between 1-6~M$_{\odot}$ (the mass range considered by Karakas). Since this group of stars is expected to produce significant amounts of C and N, it is of obvious importance to investigate PN with central star progenitors in this range. Unfortunately, these stars evolve faster along the post-AGB track, and so PN are likely to exist over a shorter time window, making finding and observing them more difficult. Nevertheless, a focus on these stars is of obvious importance for constraining LIMS models and by extension the yields of C and N by this important mass range of stars.


\acknowledgments

Support for HST program number GOÐ12600 was provided by
NASA through a grant from the Space Telescope Science
Institute, which is operated by the Association of Universities
for Research in Astronomy, Incorporated, under NASA
contract NAS5-26555. B.B. received partial support from
NSF grant AST-0808201. All authors are grateful to their home institutions for travel support. Finally, we very much appreciate the detailed and careful review of this paper by the referee.

\appendix

\section*{Appendix A: AGB Model Set Comparisons}

In section~4 we refer to AGB model sets published by \citet{buell97}, \citet{marigo01} and \citet{karakas10}. We briefly describe each of these sets below.

Buell updated the code originally developed by \citet{renzini81} and used it to compute synthetic models of thermally pulsing AGB stars for progenitor stars of masses between 1-8~M$_{\odot}$ and a metallicity of 0.02. The synthetic modeling technique was first developed by \citet{iben78}, where the effects of thermal pulsing are parameterized, obviating the need to compute a new stellar model for each time step and mass zone. Opacities were taken from \citet{rogers92} and \citet{alexander94}. Three types of mass loss were used: 1) the \citet{reimers75} relation was used during the ordinary wind phase; 2) then a rate related to the pulsation period was employed during the pulsating phase; and finally 3) once the rate in 2) exceeded {a superwind rate of $5 \times 10^{-5}$ M$_{\odot}~yr^{-1}$, $\dot{M}$ was set equal to the latter value}.

Marigo's calculations began with a Padua model star \citep{girardi00} which had been evolved from the ZAMS stage up to the thermally-pulsing AGB period. Beyond this point, then, synthetic techniques as described above were used to predict stellar yields as well as planetary nebula abundances. The Girardi et al. models employed the mass loss scheme of \citet{reimers75}, while the mass loss formalism in the later stages was taken from \citet{vw93}. Marigo's model grid included CSPN masses between 1-5~M$_{\odot}$ and Z=0.004, 0.008, and 0.02.

In contrast to the use of synthetic models by Buell and Marigo, Karakas evolved stars continuously from the ZAMS stage through the 3rd dredge-up stage and when relevant, the hot bottom burning stage. The mass loss prescription by \citet{reimers75} was employed on the first giant branch, while the scheme by \citet{vw93} was used on the AGB. After completion of the models a post-processing program was used to carry out the nucleosynthesis calculations. Karakas' models covered a range in mass of 1-6~M$_{\odot}$ and metallicities of Z=0.001, 0.004, 0.008 and 0.02.





\begin{deluxetable}{lcccccccccccc}
\tablecolumns{13}
\rotate
\tablewidth{0pc}
\tabletypesize{\scriptsize}
\tablecaption{Model Results: Line Strengths\label{lsa}}
\tablehead{
\colhead{Line ID}&\multicolumn{3}{c}{IC 2165}&\multicolumn{3}{c}{IC 3568}&\multicolumn{3}{c}{NGC 2440}&\multicolumn{3}{c}{NGC 3242}\\
\colhead{}&\colhead{Observed}&\colhead{Model}&\colhead{Model/Observed}&\colhead{Observed}&\colhead{Model}&\colhead{Model/Observed}&\colhead{Observed}&\colhead{Model}&\colhead{Model/Observed}&\colhead{Observed}&\colhead{Model}&\colhead{Model/Observed}
}

\startdata
N V $\lambda$1240	&	44.1$\pm{2.4}$	&	2.45	&	0.06	&	45.$\pm{2.8}$	&	0.24	&	0.01	&	128.$\pm{2.}$	&	37.34	&	0.29	&	2.02$\pm{0.2}$	&	1.77	&	0.88	\\
N IV] $\lambda$1486	&	47.9$\pm{2.}$	&	52.1	&	1.09	&	\nodata	&	1.69	&	\nodata	&	300.$\pm{4.}$	&	407.3	&	1.36	&	10.2$\pm{0.2}$	&	20.2	&	1.98	\\
C~IV $\lambda$1549	&	957.$\pm{39.}$	&	225.34	&	0.24	&	39.1$\pm{2.4}$ &	20.25	&	0.52	&	541.$\pm{8.}$	&	243.69	&	0.45	&	41.9$\pm{0.7}$	&	302.26	&	7.23	\\
He II $\lambda$1640	&	350.$\pm{14.}$	&	378.93	&	1.08	&	20.9$\pm{1.8}$	&	25.62	&	1.23	&	516.$\pm{7.}$	&	571.2	&	1.11	&	311.$\pm{5.}$	&	355.95	&	1.15	\\
C III] $\lambda$1906	&	424.$\pm{22.}$	&	437.25	&	1.03	&	70.6$\pm{7.}$	&	73.06	&	1.04	&	552.$\pm{14.}$	&	477.82	&	0.87	&	136.$\pm{3.}$	&	149.74	&	1.10	\\
C III]  $\lambda$1910	&	334.$\pm{19.}$	&	312.65	&	0.94	&	51.2$\pm{7.2}$	&	48.79	&	0.95	&	418.$\pm{14.}$	&	369.05	&	0.88	&	98.7$\pm{2.6}$	&	109.17	&	1.11	\\
C III] $\lambda$1909	&	758.$\pm{41.}$	&	749.9	&	0.99	&	121.8$\pm{14.2}$	&	121.85	&	1.00	&	970.$\pm{28.}$	&	846.87	&	0.87	&	234.7$\pm{5.6}$	&	258.9	&	1.10	\\
\[[O II] $\lambda$3727	&	41.2$\pm{2.5}$	&	43.13	&	1.05	&	17.4$\pm{2.5}$	&	17.58	&	1.01	&	111.$\pm{3.}$	&	97.01	&	0.87	&	11.3$\pm{2.9}$	&	11.50	&	1.02	\\
\[[Ne III] $\lambda$3869	&	90.9$\pm{2.4}$	&	89.45	&	0.98	&	77.8$\pm{1.9}$	&	80.56	&	1.04	&	97.8$\pm{2.5}$	&	93.28	&	0.95	&	94.9$\pm{0.7}$	&	95.67	&	1.01	\\
\[[O III] $\lambda$4363	&	19.5$\pm{0.6}$	&	19.30	&	0.99	&	9.23$\pm{0.5}$	&	9.15	&	0.99	&	26.9	$\pm{1.2}$&	26.80	&	1.00	&	13.1$\pm{0.2}$	&	15.80	&	1.21	\\
He II $\lambda$4686	&	54.$\pm{1.1}$	&	56.75	&	1.05	&	3.71$\pm{0.6}$	&	3.83	&	1.03	&	83.7$\pm{0.5}$	&	84.39	&	1.01	&	49.6$\pm{0.2}$	&	53.6	&	1.08	\\
H~I $\lambda$4861	&	100.	&	100.	&	1.00	&	100.	&	100.	&	1.00	&	100.	&	100.	&	1.00	&	100.	&	100.	&	1.00	\\
\[[O III] $\lambda$5007	&	1147.$\pm{10.}$	&	1233.4	&	1.08	&	1072.$\pm{11.}$	&	1070.1	&	1.00	&	1318.$\pm{3.}$	&	1406.	&	1.07	&	1191.$\pm{5.}$	&	1318.4	&	1.11	\\
He I $\lambda$5876 	&	8.76$\pm{0.6}$	&	8.12	&	0.93	&	16.6$\pm{0.6}$	&	15.571	&	0.94	&	9.89$\pm{0.6}$	&	10.12	&	1.02	&	10.3$\pm{0.3}$	&	11.53	&	1.12	\\
\[[S III] $\lambda$6312	&	4.94$\pm{0.7}$	&	1.38	&	0.28	&	\nodata	&	0.20	&	\nodata	&	1.15$\pm{0.3}$	&	1.71	&	1.49	&	\nodata	&	0.38	&	\nodata	\\
H~I $\lambda$6563	&	279.$\pm{0.}$	&	283.74	&	1.02	&	284$\pm{1.}$	&	282.79	&	1.00	&	278.$\pm{0.}$	&	288.56	&	1.04	&	282.$\pm{0.}$	&	283.52	&	1.01	\\
\[[N II] $\lambda$6584	&	34.5$\pm{0.2}$	&	34.23	&	1.00	&	2.16$\pm{0.4}$	&	2.00	&	0.93	&	580.$\pm{2.}$	&	547.28	&	0.94	&	3.4$\pm{0.3}$	&	3.41	&	1.00	\\
\[[S II] $\lambda$6716	&	2.09$\pm{0.3}$	&	0.69	&	0.33	&	\nodata	&	0.04	&	\nodata	&	4.97$\pm{0.5}$	&	1.05	&	0.21	&	0.24$\pm{0.05}$	&	0.08	&	0.32	\\
\[[S II] $\lambda$6731	&	2.73$\pm{0.3}$	&	1.20	&	0.44	&	\nodata	&	0.04	&	\nodata	&	8.55$\pm{0.5}$&	2.06&	0.24	&	0.32	$\pm{0.05}$&	0.14	&	0.43	\\
\[[Ar III] $\lambda$7135	&	9.91$\pm{0.3}$	&	9.81	&	0.99	&	6.88$\pm{0.6}$	&	6.70	&	0.97	&	19.9$\pm{0.3}$	&	20.00	&	1.01	&	6.98$\pm{0.3}$	&	6.96	&	1.00	\\
\[[S III] $\lambda$9532	&	26.$\pm{0.5}$	&	26.15	&	1.01	&	6.84$\pm{1.5}$	&	5.97	&	0.87	&	28.$\pm{0.8}$	&	28.49	&	1.02	&	10.4$\pm{0.8}$	&	9.34	&	0.90	\\
RMS\tablenotemark{a}	&	0.04	&	\nodata	&	0.05	&	0.10	&	\nodata	&	0.07	&	0.03	&	\nodata	&	0.10	&	0.08	&	\nodata	&	0.12	 \\
\enddata
\tablenotetext{a}{Root mean square excludes C~IV $\lambda$1549, N~IV] $\lambda$1486, N~V $\lambda$1240, and all S lines for reasons discussed in the text.}
\end{deluxetable}

\begin{deluxetable}{lcccccccccccc}
\tablecolumns{13}
\rotate
\tablewidth{0pc}
\tabletypesize{\scriptsize}
\tablecaption{Model Results: Line Strengths\label{lsb}}
\tablehead{
\colhead{Line ID}&\multicolumn{3}{c}{NGC 5315}&\multicolumn{3}{c}{NGC 5882}&\multicolumn{3}{c}{NGC 7662}&\multicolumn{3}{c}{PB6\tablenotemark{a}}\\
\colhead{}&\colhead{Observed}&\colhead{Model}&\colhead{Model/Observed}&\colhead{Observed}&\colhead{Model}&\colhead{Model/Observed}&\colhead{Observed}&\colhead{Model}&\colhead{Model/Observed}&\colhead{Observed}&\colhead{Model}&\colhead{Model/Observed}
}

\startdata
N V $\lambda$1240	&	6.94$\pm{5.1}$	&	0.1	&	0.02	&	9.35$\pm{1.4}$	&	0.02	&	0.002	&	3.17$\pm{0.2}$	&	8.23&	2.60	&	123.$\pm{136.}$	&	52.645&	0.43	\\
N IV] $\lambda$1486	&	$<$2.1	&	1.9	&	$>$0.92	&	\nodata	&	0.43	&	\nodata	&	15.7$\pm{0.7}$&	152.	&	9.68	&	196.$\pm{173.}$	&	133.	&	0.68	\\
C~IV $\lambda$1549	&	164.$\pm{91.}$	&	6.5	&	0.04	&	7.3$\pm{0.5}$	&	3.56	&	0.49	&	403.$\pm{17.}$	&	211.72	&	0.53	&	1288.$\pm{1100.}$	&	525.1	&	0.41	\\
He II $\lambda$1640	&	38.7$\pm{20.7}$	&	47.3	&	1.22	&	29.1$\pm{0.7}$	&	48.48	&	1.67	&	285.$\pm{12.}$	&	397.21	&	1.39	&	711.$\pm{589.}$	&	991.11	&	1.39	\\
C III] $\lambda$1906	&	23.3$\pm{2.8}$	&	32.3	&	1.39	&	12.1$\pm{4.2}$	&	13.83	&	1.14	&	204.$\pm{9.}$	&	219.39	&	1.08	&	647.$\pm{588.}$	&	554.26	&	0.86	\\
C III]  $\lambda$1910	&	35$\pm{2.8}$	&	25.6	&	0.73	&	14.8$\pm{4.2}$	&	11.99	&	0.81	&	146.$\pm{7.}$	&	154.09	&	1.06	&	306.$\pm{314.}$	&	366.15	&	1.20	\\
C III] $\lambda$1909	&	58.3$\pm{5.6}$	&	57.9	&	0.99	&	26.9$\pm{8.4}$	&	25.82	&	0.96	&	350.$\pm{16.}$	&	373.49	&	1.07	&	953.$\pm{902.}$	&	920.41	&	0.97	\\
\[[O II] $\lambda$3727	&	28.1$\pm{0.4}$	&	28.8	&	1.03	&	13.5$\pm{1.3}$	&	13.83	&	1.02	&	11.4$\pm{2.7}$	&	12.14	&	1.07	&	68.$\pm{7.}$	&	72.87	&	1.07	\\
\[[Ne III] $\lambda$3869	&	76.7$\pm{0.5}$	&	79.6	&	1.04	&	87.$\pm{1.9}$	&	83.10	&	0.96	&	92.5$\pm{1.1}$	&	104.76	&	1.13	&	110.$\pm{6.}$	&	103.61	&	0.94	\\
\[[O III] $\lambda$4363	&	4.47$\pm{0.2}$	&	4.4	&	0.99	&	6.5$\pm{0.4}$	&	6.43	&	0.99	&	18.7$\pm{0.8}$	&	26.05	&	1.39	&	18.3$\pm{2.}$	&	21.43	&	1.17	\\
He II $\lambda$4686	&	7.28$\pm{1.1}$	&	7.3	&	1.01	&	7.63$\pm{0.2}$	&	7.53	&	0.99	&	59.9$\pm{0.4}$	&	57.73	&	0.96	&	148.$\pm{24.}$	&	147.11	&	0.99	\\
H~I $\lambda$4861	&	100.	&	100.	&	1.00	&	100.	&	100.	&	1.00	&	100.	&	100.	&	1.00	&	100.	&	100.	&	1.00	\\
\[[O III] $\lambda$5007	&	791.$\pm{1.}$	&	742.9	&	0.94	&	1068.$\pm{4.}$	&	1020.8	&	0.96	&	1314.$\pm{4.}$	&	1422.	&	1.082	&	971.$\pm{49.}$	&	1058.	&	1.09	\\
He I $\lambda$5876 	&	19.4$\pm{0.2}$	&	20.1	&	1.03	&	14.$\pm{0.3}$	&	14.41	&	1.03	&	10.6$\pm{0.9}$	&	11.63	&	1.10	&	5.$\pm{0.8}$	&	5.70	&	1.14	\\
\[[S III] $\lambda$6312	&	3.39$\pm{0.03}$	&	3.6	&	1.05	&	2.09$\pm{0.3}$	&	1.51	&	0.72	&	0.97$\pm{0.3}$	&	0.97	&	1.00	&	3.5$\pm{0.5}$	&	3.65	&	1.04	\\
H~I $\lambda$6563	&	286$\pm{0.}$	&	288.1	&	1.01	&	286.$\pm{1.}$	&	286.93	&	1.00	&	280.$\pm{0.}$	&	281.5	&	1.01	&	317.$\pm{16.}$	&	291.28	&	0.92	\\
\[[N II] $\lambda$6584	&	124$\pm{0.}$	&	122.5	&	0.99	&	9.37$\pm{0.2}$	&	9.25	&	0.99	&	3.8$\pm{0.2}$	&	4.17	&	1.10	&	207.$\pm{10.}$	&	201.43	&	0.97	\\
\[[S II] $\lambda$6716	&	2.39$\pm{0.3}$	&	1.9	&	0.79	&	1.56$\pm{0.4}$	&	0.29	&	0.19	&	\nodata 	&	0.13	&	\nodata 	&	4.8$\pm{0.7}$	&	7.46	&	1.55	\\
\[[S II] $\lambda$6731	&	4.73$\pm{0.1}$	&	3.8	&	0.80	&	1.68$\pm{0.3}$	&	0.61	&	0.36	&	\nodata	&	0.22	&	\nodata	&	7.1$\pm{1.1}$	&	7.32	&	1.03	\\
\[[Ar III] $\lambda$7135	&	30.4$\pm{0.1}$	&	30.3	&	1.00	&	13.5$\pm{0.1}$	&	13.56	&	1.00	&	9.47$\pm{0.6}$	&	10.26	&	1.08	&	15.1$\pm{2.}$	&	15.35	&	1.02	\\
\[[S III] $\lambda$9532	&	140$\pm{0.}$	&	140.3	&	1.00	&	31.1$\pm{0.3}$	&	55.30	&	1.78	&	23.6$\pm{1.1}$	&	16.35	&	0.69	&	\nodata	&	57.49	&	\nodata	\\
RMS\tablenotemark{b}	&	0.07	&	\nodata	&	0.08	&	0.15	&	\nodata	&	0.20	&	0.08	&	\nodata	&	0.18	&	0.51	&	\nodata	&	0.14	\\
\enddata
\tablenotetext{a}{Strengths of lines in the optical portion of the spectrum were taken from \citet{pena98}.}
\tablenotetext{b}{Root mean square excludes C~IV $\lambda$1549, N~IV] $\lambda$1486, N~V $\lambda$1240, and all S lines for reasons discussed in the text.}
\end{deluxetable}

\begin{deluxetable}{lcccccccccc}
\rotate
\tablecolumns{11}
\tablewidth{0pc}
\tabletypesize{\scriptsize}
\tablecaption{Model Results: Abundances\tablenotemark{a}\label{abuna}}
\tablehead{
\colhead{Ratio}&\multicolumn{2}{c}{IC 2165}&\multicolumn{2}{c}{IC 3568}&\multicolumn{2}{c}{NGC 2440}&\multicolumn{2}{c}{NGC 3242}&\colhead{Solar\tablenotemark{b}}&\colhead{Orion\tablenotemark{c}}\\
\colhead{}&\colhead{Observed}&\colhead{Model}&\colhead{Observed}&\colhead{Model}&\colhead{Observed}&\colhead{Model}&\colhead{Observed}&\colhead{Model}&\colhead{}&\colhead{}
}

\startdata
He/H	&	1.06$\pm{0.1}$E-01	&	9.80E-02	&	1.18$\pm{0.1}$E-01	&	1.18E-01	&	1.27$\pm{0.1}$E-01	&	1.32E-01	&	1.15$\pm{0.1}$E-01	&	1.20E-01	&	8.50E-02	&	9.70E-02	\\
C/H	&	3.17$\pm{0.2}$E-04	&	3.30E-04	&	1.56$\pm{0.3}$E-04	&	2.04E-04	&	1.95$\pm{0.2}$E-04	&	2.59E-04	&	1.90$\pm{0.1}$E-04	&	2.42E-04	&	2.69E-04	&	3.31E-04	\\
N/H	&	8.57$\pm{0.5}$E-05	&	1.15E-04	&	1.31$\pm{0.7}$E-05	&	4.93E-05	&	4.13$\pm{0.3}$E-04	&	6.08E-04	&	8.09$\pm{0.4}$E-05	&	6.94E-05	&	6.76E-05	&	5.37E-05	\\
O/H	&	2.78$\pm{0.2}$E-04	&	2.59E-04	&	2.95$\pm{0.2}$E-04	&	3.25E-04	&	4.04$\pm{0.5}$E-04	&	2.95E-04	&	3.98$\pm{0.1}$E-04	&	3.80E-04	&	4.89E-04	&	5.37E-04	\\
Ne/H	&	5.21$\pm{0.4}$E-05	&	3.67E-05	&	5.93$\pm{0.5}$E-05	&	5.89E-05	&	6.98$\pm{1.}$E-05	&	3.83E-05	&	8.50$\pm{0.2}$E-05	&	3.22E-05	&	8.51E-05	&	1.12E-04	\\
S/H	&	1.28$\pm{0.1}$E-06	&	1.62E-06	&	5.51$\pm{1.6}$E-07	&	5.90E-07	&	1.40$\pm{0.1}$E-06	&	1.61E-06	&	9.88$\pm{1.5}$E-07	&	5.62E-06	&	1.32E-05	&	1.66E-05	\\
Ar/H	&	2.35$\pm{0.2}$E-06	&	8.67E-07	&	1.05$\pm{0.2}$E-06	&	9.33E-07	&	4.64$\pm{0.4}$E-06	&	1.53E-06	&	1.87$\pm{0.1}$E-06	&	2.59E-06	&	2.51E-06	&	4.17E-06	\\
C/O	&	1.14$\pm{0.1}$E+00	&	1.27E+00	&	5.29$\pm{1.}$E-01	&	6.28E-01	&	4.80$\pm{0.7}$E-01	&	8.78E-01	&	4.77$\pm{0.2}$E-01	&	6.37E-01	&	5.50E-01	&	5.37E-05	\\
N/O	&	3.08$\pm{0.3}$E-01	&	4.44E-01	&	4.44$\pm{2.5}$E-02	&	1.52E-01	&	1.02$\pm{0.1}$E+00	&	2.06E+00	&	2.03$\pm{0.1}$E-01	&	1.83E-01	&	1.38E-01	&	1.00E-01	\\
Ne/O	&	1.87$\pm{0.1}$E-01	&	1.42E-01	&	2.01$\pm{0.1}$E-01	&	1.81E-01	&	1.73$\pm{0.1}$E-01	&	1.30E-01	&	2.13$\pm{0.1}$E-01	&	8.47E-02	&	1.74E-01	&	2.09E-01	\\
S/O	&	4.60$\pm{0.3}$E-03	&	6.25E-03	&	1.87$\pm{0.6}$E-03	&	1.82E-03	&	3.47$\pm{0.4}$E-03	&	5.46E-03	&	2.48$\pm{0.4}$E-03	&	1.48E-02	&	2.70E-02	&	3.09E-02	\\
Ar/O	&	8.45$\pm{0.5}$E-03	&	3.35E-03	&	3.57$\pm{0.6}$E-03	&	2.87E-03	&	1.15$\pm{0.1}$E-02	&	5.19E-03	&	4.70$\pm{0.2}$E-03	&	6.82E-03	&	5.13E-03	&	7.77E-03	\\
\enddata
\tablenotetext{a}{Observed abundances and uncertainties are taken directly from \citet{dufour15}.}
\tablenotetext{b}{Asplund et al. (2009)}
\tablenotetext{c}{Esteban et al. (2004)}
\end{deluxetable}

\begin{deluxetable}{lcccccccccc}
\tablecolumns{11}
\rotate
\tablewidth{0pc}
\tabletypesize{\scriptsize}
\tablecaption{Model Results: Abundances\tablenotemark{a}\label{abunb}}
\tablehead{
\colhead{Ratio}&\multicolumn{2}{c}{NGC 5315}&\multicolumn{2}{c}{NGC 5882}&\multicolumn{2}{c}{NGC 7662}&\multicolumn{2}{c}{PB6}&\colhead{Solar\tablenotemark{b}}&\colhead{Orion\tablenotemark{c}}\\
\colhead{}&\colhead{Observed}&\colhead{Model}&\colhead{Observed}&\colhead{Model}&\colhead{Observed}&\colhead{Model}&\colhead{Observed}&\colhead{Model}&\colhead{}&\colhead{}
}

\startdata
He/H	&	1.32$\pm{0.1}$E-01	&	1.70E-01	&	1.10$\pm{0.1}$E-01	&	9.90E-02	&	1.22$\pm{0.1}$E-01	&	1.20E-01	&	1.80$\pm{0.1}$E-01	&	1.80E-01	&	8.50E-02	&	9.70E-02	\\
C/H	&	2.34$\pm{0.3}$E-04	&	1.00E-05	&	7.86$\pm{2.1}$E-05	&	9.32E-05	&	2.36$\pm{0.2}$E-04	&	1.35E-04	&	8.32$\pm{1.2}$E-04	&	2.25E-04	&	2.69E-04	&	3.31E-04	\\
N/H	&	1.49$\pm{0.3}$E-04	&	5.33E-04	&	5.76$\pm{1.4}$E-05	&	1.01E-04	&	4.95$\pm{0.5}$E-05	&	9.36E-05	&	4.17$\pm{0.5}$E-04	&	3.38E-04	&	6.76E-05	&	5.37E-05	\\
O/H	&	3.65$\pm{0.2}$E-04	&	1.04E-04	&	4.42$\pm{0.3}$E-04	&	4.47E-04	&	3.54$\pm{0.2}$E-04	&	2.66E-04	&	3.20$\pm{0.5}$E-04	&	1.94E-04	&	4.89E-04	&	5.37E-04	\\
Ne/H	&	9.74$\pm{0.5}$E-05	&	3.35E-05	&	1.01$\pm{0.1}$E-04	&	9.10E-05	&	6.34$\pm{0.4}$E-05	&	3.79E-05	&	8.13$\pm{0.5}$E-05	&	3.51E-05	&	8.51E-05	&	1.12E-04	\\
S/H	&	1.47$\pm{0.1}$E-05	&	2.28E-05	&	5.37$\pm{0.7}$E-06	&	5.37E-06	&	2.03$\pm{0.4}$E-06	&	1.57E-06	&	1.29$\pm{0.5}$E-05	&	3.09E-06	&	1.32E-05	&	1.66E-05	\\
Ar/H	&	3.34$\pm{0.1}$E-06	&	3.51E-06	&	2.28$\pm{0.1}$E-06	&	1.99E-06	&	2.06$\pm{0.2}$E-06	&	1.57E-06	&	3.89$\pm{0.5}$E-06	&	2.55E-06	&	2.51E-06	&	4.17E-06	\\
C/O	&	6.40$\pm{0.8}$E-01	&	9.62E-02	&	1.78$\pm{0.5}$E-01	&	2.09E-01	&	6.70$\pm{0.5}$E-01	&	5.08E-01	&	2.60$\pm{1.1}$E+00	&	1.16E+00	&	5.50E-01	&	5.37E-05	\\
N/O	&	4.10$\pm{0.9}$E-01	&	5.13E+00	&	1.30$\pm{0.3}$E-01	&	2.26E-01	&	1.40$\pm{0.2}$E-01	&	3.52E-01	&	1.30$\pm{0.2}$E+00	&	1.74E+00	&	1.38E-01	&	1.00E-01	\\
Ne/O	&	2.67$\pm{0.1}$E-01	&	3.22E-01	&	2.29$\pm{0.1}$E-01	&	2.04E-01	&	1.79$\pm{0.1}$E-01	&	1.42E-01	&	2.54$\pm{0.1}$E-01	&	1.81E-01	&	1.74E-01	&	2.09E-01	\\
S/O	&	4.03$\pm{0.4}$E-02	&	2.19E-01	&	1.22$\pm{0.1}$E-02	&	1.20E-02	&	5.72$\pm{1.2}$E-03	&	5.90E-03	&	4.03$\pm{0.1}$E-02	&	1.59E-02	&	2.70E-02	&	3.09E-02	\\
Ar/O	&	9.16$\pm{0.2}$E-03	&	3.38E-02	&	5.15$\pm{0.3}$E-03	&	4.45E-03	&	5.81$\pm{0.4}$E-03	&	5.90E-03	&	1.22$\pm{0.1}$E-02	&	1.31E-02	&	5.13E-03	&	7.77E-03	\\
\enddata
\tablenotetext{a}{Observed abundances and uncertainties for all objects but PB6 are taken directly from \citet{dufour15}. Those for PB6 are taken from \citet{pena98}.}
\tablenotetext{b}{Asplund et al. (2009)}
\tablenotetext{c}{Esteban et al. (2004)}
\end{deluxetable}

\clearpage

\begin{deluxetable}{lcccccccc}
\tablecolumns{9}
\rotate
\tablewidth{0pc}
\tabletypesize{\scriptsize}
\tablecaption{Model Results: Parameters}
\tablehead{
\colhead{Parameter}&\multicolumn{2}{c}{IC 2165}&\multicolumn{2}{c}{IC 3568}&\multicolumn{2}{c}{NGC 2440}&\multicolumn{2}{c}{NGC 3242}\\
\colhead{}&\colhead{Observed}&\colhead{Model}&\colhead{Observed}&\colhead{Model}&\colhead{Observed}&\colhead{Model}&\colhead{Observed}&\colhead{Model}
}

\startdata
\cutinhead{Stellar}
T$_{eff} (kK)$\tablenotemark{a}	&	115$\pm$3	&	110	&	51$\pm$1	&	51	&	166$\pm$40	&	198	&	85$\pm$7	&	89	\\
log(L*/L$_{\odot}$)\tablenotemark{a}	&	3.87$\pm$0.10	&	3.16	&	3.98$\pm$0.14	&	3.88	&	3.63$\pm$0.36	&	3.10	&	3.55$\pm$0.03	&	3.64	\\
R*(R$_{\odot}$)\tablenotemark{b}	&	0.21	&	0.10	&	1.25	&	1.12	&	0.08	&	0.03	&	0.27	&	0.28	\\
M(M$_{\odot}$)\tablenotemark{c}	&	2.3	&	0.57/1.0	&	2.5	&	0.63/2.0	&	2.1	&	0.73/3.2	&	1.2	&	0.60/1.5	\\
\cutinhead{Nebular}																	\\
log(L$_{H\beta}$) (erg/s)\tablenotemark{d}	&	34.0$\pm{0.29}$	&	34.0	&	34.1$\pm{0.06}$	&	34.1	&	34.2$\pm{0.33}$	&	34.2	&	34.4$\pm{0.04}$	&	34.3	\\
N$_e$ (C III]; cm$^{-3}$)\tablenotemark{e}	 	&	8000$\pm$2500 &	3000	&	3600$\pm$6400	&	900	&	6400$\pm$1800 & 	5000	&	3800$\pm$1000	&	4000	\\
N$_e$ ([S II]; cm$^{-3}$)\tablenotemark{e}	 	&	1600$\pm$1100 &	3500	&	\nodata	&	800	&	4600$\pm$2400 & 	6000	&	\nodata	&	500	\\
N$_e$ (H${\beta}$; cm$^{-3}$) 	&	3300	&	3800	&	700	&	800	&	900	&	7600	&	1200	&	4400	\\
N$_H$ (model input; cm$^{-3}$)\tablenotemark{f} 	&	\nodata	&	3300	&	\nodata	&	700	&	\nodata	&	6300	&	\nodata	&	3800	\\
Filling factor	&	\nodata	&	0.2	&	\nodata	&	1.0	&	\nodata	&	0.5	&	\nodata	&	0.2	\\
T$_e$([O~III]; K)\tablenotemark{e}	&	14,200$\pm$200 &	13,100	&	11,000$\pm$200	&	11,100	&	15,300$\pm$300	&	14,500	&	12,000$\pm$100	&	11,700	\\
Radius (pc)\tablenotemark{g}	&	0.04	&	0.04	&	0.12	&	0.08	&	0.11	&	0.01	&	0.10\	&	0.05	\\
\enddata
\tablenotetext{a}{Observed values are averages of values reported by \citet{shaw89,zhang93,frew08} and listed separately in Table~\ref{paramc}; uncertainties are the RMS of the offsets from the averages}
\tablenotetext{b}{Computed from $T_{eff}$ and log(L/L$_{\odot}$) using the Stefan-Boltzmann law}
\tablenotetext{c}{Final/Initial model values presented here were determined in \S~\ref{properties}} 
\tablenotetext{d}{Observed values are derived from fluxes reported by \citet{cahn92} and using distances from \citet{cahn92,zhang93,kwitter06}}
\tablenotetext{e}{Observed values are those reported by \citet{dufour15}; model values for C~III] and [S II] densities were computed from the predicted $\lambda$1906/$\lambda$1910 and $\lambda$6716/$\lambda$6731 ratios, respectively, using information in \citet{osterbrock06}}
\tablenotetext{f}{Total H density used as input for the final model}
\tablenotetext{g}{Observed values are taken directly from \citet{cahn92}}
\label{parama}
\end{deluxetable}

\begin{deluxetable}{lcccccccc}
\tablecolumns{9}
\rotate
\tablewidth{0pc}
\tabletypesize{\scriptsize}
\tablecaption{Model Results: Parameters}
\tablehead{
\colhead{Parameter}&\multicolumn{2}{c}{NGC 5315}&\multicolumn{2}{c}{NGC 5882}&\multicolumn{2}{c}{NGC 7662}&\multicolumn{2}{c}{PB6}\\
\colhead{}&\colhead{Observed}&\colhead{Model}&\colhead{Observed}&\colhead{Model}&\colhead{Observed}&\colhead{Model}&\colhead{Observed}&\colhead{Model}
}

\startdata
\cutinhead{Stellar}
T$_{eff} (kK)$\tablenotemark{a}	&	60$\pm$1	&	70	&	70$\pm$2	&	70	&	107$\pm$7	&	95	&	165$\pm$25	&	150	\\
log(L*/L$_{\odot}$)\tablenotemark{a}	&	3.95$\pm$0.10	&	3.50	&	3.72$\pm$0.16&	3.45	&	3.73$\pm$0.14 &	3.42	&	3.43$\pm$0.10	&	3.20	\\
R*(R$_{\odot}$)\tablenotemark{b}	&	0.87	&	0.38	&	0.49	&	0.36	&	0.21&	0.19	&	0.06	&	0.06	\\
M(M$_{\odot}$)\tablenotemark{c}	&	2.5	&	0.57/1.1	&	1.5	&	0.56/1.0	&	1.8	&	0.57/1.0	&	1.9	&	0.62/1.8	\\
\cutinhead{Nebular}																	
log(L$_{H\beta}$) (erg/s)\tablenotemark{d}	&	34.4$\pm{0.05}$	&	34.5	&	34.2$\pm{0.05}$\ &	34.4	&	34.2$\pm{0.17}$	&	34.1	&	33.5$\pm{0.02}$	&	33.5	\\
N$_e$ (C III]; cm$^{-3}$)\tablenotemark{e} 	&	46,600$\pm$11,000	&	5000	&	29,000$\pm$29,000	&	10,000	&	3300$\pm$1000&	2000	&	\nodata	&	900	\\
N$_e$ ([S II]; cm$^{-3}$)\tablenotemark{e} 	&	8800$\pm$2600	&	7200	&	800$\pm$800	&	10,200	&	\nodata &	3200	&	2100$\pm{900}$	&	1600	\\
N$_e$ (H${\beta}$; cm$^{-3}$) 	&	2600	&	7200	&	2400	&	11,100	&	1300	&	3200	&	400	&	500	\\
N$_H$ (model input; cm$^{-3}$)\tablenotemark{f} 	&	\nodata	&	6300	&	\nodata	&	10,000	&	\nodata	&	2700	&	\nodata	&	1600	\\
Filliing factor	&	\nodata	&	0.09	&	\nodata	&	0.5	&	\nodata	&	1	&	\nodata	&	0.04	\\
T$_e$([O~III]; K)\tablenotemark{e}	&	9600$\pm$125	&	9400	&	9900$\pm$200	&	9700	&	13,200$\pm$200	&	14,100	&	14,800	&	15,000	\\
Radius (pc)\tablenotemark{g}	&	0.02	&	0.05	&	0.06	&	0.01	&	0.04	&	0.02	&	0.12	&	0.31	\\
\enddata
\tablenotetext{a}{Observed values are averages of values reported by \citet{shaw89,zhang93,frew08} and listed separately in Table~\ref{paramc}; uncertainties are the RMS of the offsets from the averages. T$_{eff}$ and log(L/L$_{\odot}$) for PB6 are taken directly from \citet{keller14}, although the uncertainties represent our estimates.}
\tablenotetext{b}{Computed from $T_{eff}$ and log(L/L$_{\odot}$) using the Stefan-Boltzmann law}
\tablenotetext{c}{Final/Initial model values presented here were determined in \S~\ref{properties}} 
\tablenotetext{d}{Observed values are derived from fluxes reported by \citet{cahn92} and using distances from \citet{cahn92,zhang93,kwitter06}}
\tablenotetext{e}{Observed values are those reported by \citet{dufour15} and \citet[PB6 S II density only]{garcia09}; model values for C~III] and [S II] densities were computed from the predicted $\lambda$1906/$\lambda$1910 and $\lambda$6716/$\lambda$6731 ratios, respectively, using information in \citet{osterbrock06}}
\tablenotetext{f}{Total H density used as input for the final model}
\tablenotetext{g}{Observed values are taken directly from \citet{cahn92}}
\label{paramb}
\end{deluxetable}

\begin{deluxetable}{lccccc}
\tablecolumns{6}
\tablewidth{0pc}
\tabletypesize{\scriptsize}
\tablecaption{Stellar Parameters Comparison}
\tablehead{
\colhead{Object} & \colhead{Frew\tablenotemark{a}} & \colhead{Zhang\tablenotemark{b} }& \colhead{Shaw\tablenotemark{c}} & \colhead{Average} &\colhead{Model} 
}
\startdata
\cutinhead{Effective Temperature (kK)}
IC 2165 & \nodata&112 &118 &115 & 110 \\
IC 3568 &\nodata  &51 &52 &51 &51 \\
NGC 2440 &208  &179 &112 &166 &198 \\
NGC 3242 &89  &75 &90 &85  &89 \\
NGC 5315 &\nodata  &60 &61 &60 &77 \\
NGC 5882 &68  &73 &70 &70 &70 \\
NGC 7662 &111  &97 &113 &107 &95 \\
PB6 & \nodata & \nodata & \nodata & 165\tablenotemark{d} & 150 \\
\cutinhead{log Luminosity (L$_{\odot}$)}
IC 2165 & \nodata&3.95 &3.76 &3.87 &3.16 \\
IC 3568 &\nodata  &3.78 &4.12 &3.98 &3.88 \\
NGC 2440 &3.32  &3.51 &3.88 &3.63 &3.10 \\
NGC 3242 &3.54 &3.53 &3.59  &3.55&3.64 \\
NGC 5315 &\nodata  &3.95 &$<$4.25 &3.95 &3.51 \\
NGC 5882 &3.52  &3.85 &3.72 &3.72 &3.45 \\
NGC 7662 &3.42  &3.76 &3.89 &3.73 &3.42 \\
PB6 & \nodata & \nodata & \nodata & 3.43\tablenotemark{d} &3.20 
\enddata
\tablenotetext{a}{\citet{frew08}}
\tablenotetext{b}{\citet{zhang93}}
\tablenotetext{c}{\citet{shaw85,shaw89}}
\tablenotetext{d}{\citet{keller14}}
\label{paramc}
\end{deluxetable}

\begin{deluxetable}{lcccccccc}
\tablecolumns{9}
\tablewidth{0pc}
\tabletypesize{\scriptsize}
\tablecaption{Comparison of Observed and Predicted Carbon and Nitrogen Abundances\tablenotemark{a}}
\tablehead{
\multicolumn{2}{c}{}&\colhead{O/H}&\colhead{C/H}&\colhead{C/H}&\colhead{C/H}&\colhead{N/H}&\colhead{N/H}&\colhead{N/H} \\
\colhead{Object} & \colhead{Mass} & \colhead{Observed}&\colhead{Observed}& \colhead{Predicted}&\colhead{Pred/Obs}&\colhead{Observed}& \colhead{Predicted}&\colhead{Pred/Obs} 
}
\startdata
IC 2165			&	1.0	&	2.8$\pm{0.2}$	&	3.2$\pm{0.2}$	&	1.0	&	0.3	&	0.9$\pm{0.1}$	&	0.6	&	0.7	\\
NGC 5882	&	1.0	&	4.4$\pm{0.3}$	&	0.8$\pm{0.2}$	&	1.7	&	2.1	&	0.6$\pm{0.1}$	&	0.9	&	1.6	\\
NGC 7662	&	1.0	&	3.5$\pm{0.2}$	&	2.4$\pm{0.2}$	&	0.8	&	0.3	&	0.5$\pm{0.1}$	&	0.7	&	1.5	\\
NGC 5315	&	1.1	&	3.7$\pm{0.2}$	&	2.3$\pm{0.3}$	&	1.3	&	0.6	&	1.5$\pm{0.3}$	&	0.8	&	0.5	\\
NGC 3242	&	1.5	&	4.0$\pm{0.1}$	&	1.9$\pm{0.1}$	&	1.5	&	0.8	&	0.8$\pm{0.1}$	&	1.0	&	1.3	\\
PB6				&	1.8	&	3.2$\pm{0.5}$	&	8.3$\pm{1.2}$	&	4.9	&	0.6	&	4.2$\pm{0.5}$	&	0.9	&	0.2	\\
IC 3568			&	 2.0	&	3.0$\pm{0.2}$	&	1.6$\pm{0.3}$	&	8.0	&	5.1	&	0.1$\pm{0.1}$	&	0.8	&	6.3	\\
NGC 2440	&	3.2	&	4.0$\pm{0.5}$	&	2.0$\pm{0.2}$	&	17.1	&	8.8	&	4.1$\pm{0.3}$	&	1.4	&	0.3	\\
\enddata
\tablenotetext{a}{Observed abundances are taken from Paper~I, while predicted abundances are from \citet{karakas10}. Abundance ratios listed in columns 3, 4, 5, 7, and 8 are in units of $10^{-4}$.}
\label{comparisons}
\end{deluxetable}


\begin{thebibliography} {}

\bibitem[Aldrovandi \& P{\'e}quignot (1973)]{aldrovandi73}Aldrovandi, S.M.V., and P{\'e}quignot, D. 1973, \aap, 25,137

\bibitem[Alexander \& Ferguson (1994)]{alexander94}Alexander, D.R., and Ferguson, J.W. 1994, \apj, 437, 879

\bibitem[Anders \& Grevesse (1989)]{anders89}Anders, E., and Grevesse, N. 1989, Geochim. Cosmochim. Acta, 53, 197

\bibitem[Asplund et al. (2009)]{asplund09}Asplund, M., Grevesse, N., Sauval, A.J. and Scott, P. 2009, \araa, 47, 481

\bibitem[Badnell et al. (2015)]{badnell15}Badnell, N.R., Ferland, G.J., Gorczyca, T.W., et al. 2015, \apj, 840, 100

\bibitem[Bassgen et al. (1995)]{bassgen95} Bassgen, M., Diesch, C., and Grewing, M. 1995, \aap, 297, 828

\bibitem[Blair et al. (1981)]{blair81} Blair, W.P., Kirshner, R.P., and Chevalier, R.A. 1981, \apj, 247, 879

\bibitem[Bl\"{o}cker (1995)]{blocker95}Bl\"{o}cker, T. 1995, \aap, 299, 755

\bibitem[Bohigas et al. (2013)]{bohigas13} Rev Mex A\&Af 2013, 49, 227

\bibitem[Bohlin et al. (1978)]{bohlin78}Bohlin, R.C., Harrington, J.P., and Stecher, T.P. 1978, \apj, 219, 575

\bibitem[Buell (1997)]{buell97}Buell, J.F. 1997, PhD thesis, University of Oklahoma

\bibitem[Cahn et al. (1992)]{cahn92}Cahn, J.H., Kaler, J.B., and Stanghellini, L. 1992, \aaps, 94, 399

\bibitem[Carigi et al. (2005)]{carigi05}Carigi, L., Peimbert, M., Esteban, C., \& Garc\'{i}a-Rojas, J. 2005, \apj, 623, 213

\bibitem[Chiappini et al. (1997)]{chiappini97}Chiappini, C., Matteucci, F., \& Gratton, R. 1997, \apj, 477, 765

\bibitem[Chiappini et al. (2003)]{chiappini03}Chiappini, C., Romano, D., \& Matteucci, F. 2003, \mnras, 339, 63

\bibitem[Dopita (1977)]{dopita77}Dopita, M.A. 1977, \apjs, 33, 437

\bibitem[Dufour (1991)]{dufour91}Dufour, R.J. 1991, \pasp, 103, 857

\bibitem[Dufour et al. (2015)]{dufour15}Dufour, R.J., Kwitter, K.B., Shaw, R.A., Henry, R.B.C., Balick, B., and Corradi, R.L.M. 2015, \apj, 803, 23 (Paper~I)

\bibitem[Esteban et al. (2004)]{esteban04}Esteban, C., Peimbert, M., Garc{\'i}a-Rojas, et al. (2004), \mnras, 355, 229

\bibitem[Ferland et al. (2013)]{ferland13}Ferland, G. J., Porter, R. L., van Hoof, P. A. M., et al. 2013, Rev. Mexicana Astron.
Astros., 49, 137

\bibitem[Frew (2008)]{frew08} Frew, D. 2008, PhD thesis, Macquarie University

\bibitem[Frew \& Parker (2010)]{frew10}Frew, D.J., and Parker, Q.A. 2010, \pasa, 27, 129

\bibitem[Garc\'{i}a-Rojas et al. (2009)]{garcia09}Garc\'{i}a-Rojas, J., Pe\~{n}a, M., and Peimbert, A. 2009, \aap, 496, 139

\bibitem[Gavil\'{a}n et al. (2005)]{gavilan05}Gavil\'{a}n, M., Buell, J.F., \& Moll\'{a}, M. 2005, \aap, 432, 861

\bibitem[Gavil\'{a}n et al. (2006)]{gavilan06}Gavil\'{a}n, M., Moll\'{a}, M., \& Buell, J.F. 2006, \aap, 450, 509

\bibitem[Gesicki \& Zijlstra (2007)]{gesicki07}Gesicki, K., and Zijlstra, A.A. 2007, \aap, 467, L29

\bibitem[Girardi et al. (2000)]{girardi00}Girardi, L., Bressan, A., Bertelli, G., and Chiosi, C. 2000, \aaps, 354, 169

\bibitem[Groenewegen \& Marigo (2004)]{groenewegen04}Groenewegen, M.A.T., and Marigo, P. 2004, in Asymptotic Giant Branch Stars, H.J. Habig and H. Olofsson, eds., (New York: Springer), pg. 105 

\bibitem[Harrington \& Feibelman (1983)]{harrington83}Harrington, J.P., and Feibelman, W.A. 1983, \apj, 265, 258

\bibitem[Harrington et al. (1982)]{harrington82}Harrington, J.P., Seaton, M.J., Adams, S., and Lutz, J.H. 1982, \mnras, 199, 517

\bibitem[Henry et al. (1996)]{hkh96}Henry, R.B.C., Kwitter, K.B., and Howard, J.W. 1996, \apj, 458, 215

\bibitem[Henry et al. (2000a)]{henry00a}Henry, R.B.C., Kwitter, K.B., \& Bates, J.A. 2000a, \apj, 531, 928

\bibitem[Henry et al. (2000b)]{henry00b}Henry, R.B.C., Edmunds, M.G., \& K\H{o}ppen, J. 2000, \apj, 541, 660

\bibitem[Henry et al. (2004)]{hkb04}Henry, R.B.C., Kwitter, K.B., and Balick, B. 2004, \aj, 127, 2284

\bibitem[Henry et al. (2012)]{henry12}Henry, R.B.C., Speck, A., Karakas, A.I., Ferland, G.J., \& Maguire, M. 2012, \apj, 749, 61

\bibitem[Herwig (2005)]{herwig05}Herwig, F. 2005, \araa,  43, 435

\bibitem[Iben \& Truran (1978)]{iben78}Iben, I., and Truran, J.W. 1978, \apj, 220, 980


\bibitem[Karakas et al. (2002)]{karakas02}Karakas, A.I., Lattanzio, J.C., and Pols, O.R. 2002, \pasp, 19, 515

\bibitem[Karakas et al. (2009)]{karakas09}Karakas, A.I., van Raai, M.A., Lugaro, M., Sterling, N.C., and Dinerstein, H.L. 2009, \apj, 690, 1130

\bibitem[Karakas (2010)]{karakas10}Karakas, A.I. 2010, \mnras, 403, 1413

\bibitem[Karakas \& Lattanzio (2014)]{karakas14}Karakas, A.I., and Lattanzio, J.C. 2014,  \pasa, 31, 30

\bibitem[Keller et al. (2014)]{keller14}Keller, G., Bianchi, L., and Maciel, W.J. 2014, \mnras, 442, 1379

\bibitem[Kingsburgh \& Barlow (1994)]{kb94}Kingsburgh, R.L., and Barlow, M.J. 1994, \mnras, 271, 257

\bibitem[Kwitter \& Henry (1996)]{kh96}Kwitter, K.B., and Henry, R.B.C. 1996, \apj, 473, 304


\bibitem[Kwitter \& Henry (2006)]{kwitter06}Kwitter, K.B., and Henry, R.B.C. 2006, Gallery of Planetary Nebula Spectra, http://web.williams.edu/Astronomy/research/PN/nebulae/

\bibitem[Lattanzio \& Wood (2004)]{lattanzio04}Lattanzio, J.C., and Wood, P.R. 2004, in Asymptotic Giant Branch Stars, H.J. Habig and H. Olofsson, eds., (New York: Springer), pg. 23 

\bibitem[Liebert et al. (2005)]{liebert05}Liebert, J., Bergeron, P., and Holberg, J.B. 2005, \apjs, 156, 47

\bibitem[Marigo (2001)]{marigo01}Marigo, P. 2001, \aap, 370, 194

\bibitem[Marigo et al. (2003)]{marigo03}Marigo, P., Bernard-Salas, J., Pottasch, S.R., Tielens, A.G.G.M., and Wesselius, P.R. 2003, \aap, 409 619

\bibitem[Martin (1981)]{martin81}Martin, W. 1981, \aap, 98, 328

\bibitem[M\'{e}ndez et al. (1988)]{mendez88}M\'{e}ndez, R.H., Kudritzki, R.P., Herrero, A., Husfeld, D., \& Groth, H.G. 1988, \aap, 190, 113

\bibitem[Miller et al. (2015)]{miller15}Miller, T.R. 2015, in preparation

\bibitem[Osterbrock \& Ferland (2006)]{osterbrock06}Osterbrock, D.E., and Ferland, G.J. 2006, in Astrophysics of Gasious Nebulae and Active Galactic Nuclei, (Sausalito: University Science Books)

\bibitem[Pe\~{n}a et al. (1998)]{pena98}Pe\~{n}a, M., Stasi\'{n}ska, G., Esteban, C., Koesterke, L., Medina, S., \& Kingsburgh, R. 1998, \aap, 337, 866

\bibitem[P\'{e}quignot (1980)]{pequignot80}P\'{e}quignot, D. 1980, \aap, 83, 52

\bibitem[Rauch (2003)]{rauch03}Rauch, T. 2003, \aap, 320, 237 

\bibitem[Reimers (1975)]{reimers75}Reimers, D., 1975, in Baschek, B., Kegel, W.H., and Traving, G., eds, Problems in Stellar Atmospheres and Envelopes, (New York: Springer), p. 229

\bibitem[Renzini \& Voli (1981)]{renzini81} Renzini, A., \& Voli, M. 1981, \aap, 94, 175

\bibitem[Rogers \& Iglesias (1992)]{rogers92}Rogers, F.J., and Iglesias, C.A. 1992, \apjs, 79, 507


\bibitem[Salpeter (1955)]{salpeter55}Salpeter, E.E. 1955, \apj, 121, 161

\bibitem[Sch\"{o}nberner (1983)]{schonberner83}Sch\"{o}nberner D. 1983, \apj, 272, 708

\bibitem[Shaw \& Kaler (1985)]{shaw85}Shaw, R.A., and Kaler, J.B. 1985, \apj, 295, 537

\bibitem[Shaw \& Kaler (1989)]{shaw89} Shaw, R.A., and Kaler, J.B. 1989, \apjs, 69, 495

\bibitem[Shields et al. (1981)]{shields81} Shields, G.A., Aller, L.H., Keyes, C.D., and Czyzak, S.J. 1981, \apj, 248, 569

\bibitem[Shull and McKee (1979)]{shull79} Shull, J.M, and McKee, C.F. 1979, \apj, 227, 131




\bibitem[Vassiliadis \& Wood (1993)]{vw93}Vassiliadis, E., and Wood, P.R. 1993, \apj, 413,641

\bibitem[Vassiliadis \& Wood (1994)]{vw94}Vassiliadis, E., and Wood, P.R. 1994, \apjs, 92, 125

\bibitem[Wright et al. (2011)]{wright11}Wright, N.J., Barlow, M.J., Ercolano, B., and Rauch, T. 2011, \mnras, 418, 370


\bibitem[Zhang \& Kwok (1993)]{zhang93}Zhang, C.Y., and Kwok, S. 1993, \apjs, 88, 137

\end{thebibliography}
\end{document}